\documentclass[11pt, onecolumn, journal]{IEEEtran}
\usepackage{amsmath,amsfonts}
\usepackage{algorithmic}
\usepackage{algorithm}
\usepackage{array}
\usepackage[caption=false,font=normalsize,labelfont=sf,textfont=sf]{subfig}
\usepackage{textcomp}
\usepackage{stfloats}
\usepackage{url}
\usepackage{verbatim}
\usepackage{graphicx}
\usepackage{cite}
\usepackage{xcolor}
\usepackage{pifont}
\usepackage{caption}
\usepackage{tikz}
\usetikzlibrary{arrows.meta,fit,positioning}
\usepackage{xurl}
\usepackage{multirow}
\usepackage{makecell}

\usepackage[hidelinks]{hyperref}
\definecolor{introcolor}{RGB}{214,234,248}      
\definecolor{archcolor}{RGB}{232,218,239}       

\newcommand{\cmark}{\textcolor{green!60!black}{\ding{51}}}  
\newcommand{\xmark}{\textcolor{red!70!black}{\ding{55}}}    
\DeclareRobustCommand{\pmark}{%
\begin{tikzpicture}[baseline=-0.6ex]
  \draw[orange] (0,0) circle (0.6ex);
  \clip (0,0) circle (0.6ex);
  \fill[orange!80!orange] (-0.6ex,-0.6ex) rectangle (0,0.6ex);
\end{tikzpicture}%
} 

\begin{document}

\title{Synthetic Biological Intelligence: System-Level Abstractions and Adaptive Bio-Digital Interaction}

\author{Martín Schottlender, Pengjie Zhou, Veronika Volkova, Fatima Rani, Ruifeng~Zheng, Juan A. Cabrera, Frank H.\,P. Fitzek, and Pit Hofmann
\vspace{-0.3cm}
\thanks{M.~Schottlender, P.~Zhou, V.~Volkova,  F.~Rani, R.~Zheng, J. A.~Cabrera, F. H.\,P.~Fitzek, and P.~Hofmann are with the Deutsche Telekom Chair of Communication Networks, Dresden University of Technology, Germany; J. Cabrera, F.~Fitzek, and P.~Hofmann are also with the Centre for Tactile Internet with Human-in-the-Loop (CeTI), Dresden, Germany, Emails: \{martin.schottlender, pengjie.zhou, veronika.volkova, fatima.rani, ruifeng.zheng, juan.cabrera, pit.hofmann,  frank.fitzek\}@tu-dresden.de.
}
}


\maketitle

\begin{abstract}
Concurrent advances across fields such as organoid technology, Microelectrode Arrays (MEAs), neuromorphic computing, and machine learning have given rise to a groundbreaking research paradigm: \textit{Synthetic Biological Intelligence} (SBI). SBI refers to engineered systems in which living Biological Neural Networks (BNNs) are interfaced with hardware and software to perform task-oriented information processing in a closed loop. This cutting-edge technology, while still in its infancy, has the potential to deliver highly efficient performance across both computing capabilities and energy consumption. The early stage of this field underscores the need for reliable multi-scale and cross-domain interaction interfaces to support applications in robotics, biomedicine, signal processing, and neuroscience research. The hitherto lack of commercially available SBI platforms has slowed the development, as the conditions to produce a testbed are expensive and cumbersome. The introduction of standardized, platform- and cloud-integrated BNNs has been a crucial catalyst for the scientific community, improving the accessibility of SBI and leading the way to further developments. In this survey, we summarize the innovations that contributed to the emergence of SBI and the first testbed interfaces that enabled its embodiment. This work reframes SBI as a bio-digital interaction system and introduces a unified protocol across encoding, decoding, system engineering, and benchmarking.
\end{abstract}


\section{Introduction} 
\label{sec:introduction}

\IEEEPARstart{T}{he} rapid expansion of Artificial Intelligence (AI) has profoundly impacted society, and its resource footprint highlights the need for more energy-efficient computational paradigms. AI-driven infrastructure is projected to significantly increase energy demands, nearly doubling its share in global consumption between 2024 and 2030~\cite{EnergyAIAnalysis2025}. These trends highlight fundamental limitations in conventional silicon-based computing and motivate the exploration of alternatives capable of delivering high computational performance at significantly lower energy costs. 

Biologically-inspired computing is an emerging paradigm in which computational principles from biological systems are leveraged to enable more efficient computation, not only in terms of energy consumption but also in training time and scalability \cite{leviDevelopmentApplicationsBiomimetic2018}. Millions of years of evolution have produced one of the most energy- and computationally efficient systems known: the human brain. It can perform highly complex tasks while learning from a small amount of data, and do so with astonishing energy efficiency \cite{milinkovicBiologicalArtificialConsciousness2026}. The world's most powerful supercomputer at the time of writing (February, 2026), \textit{El Capitan}, has an estimated computing power of $1.742\  \text{exaflops}$, while the human brain is estimated to be on the order of $1\  \text{exaflop}$ \cite{HewlettPackardEnterprise}. However, this improvement in performance is not reflected in energy consumption, which is approximately $29.6\ \mathrm{MW}$, a far cry from the only $20\ \mathrm{W}$ that the human brain consumes \cite{smirnovaOrganoidIntelligenceOI2023}.

The current state of the art in computing is based on silicon-based digital architectures. These have enabled outstanding performance advances but are increasingly constrained by energy consumption, architectural scalability, and computational limitations, such as in deep learning for inverse problems~\cite{bocheLimitationsDeepLearning2023}. Alternatives to digital computing, also known as beyond-digital computing, have been widely discussed. These differ from classical binary abstractions and Boolean logic, and harness the properties of physical processes for information processing. Examples include quantum computing, analog computing, and biologically-inspired computing.  In particular, the paradigm of biologically-inspired computing has been under development for the past few years, with Spiking Neural Networks (SNNs), neuromorphic computing, and biological computing leading the way toward a wider availability of these technologies \cite{seokNeumannArchitectureBrainInspired2024,schmidgallBraininspiredLearningArtificial2023}. A more detailed explanation of the different terminologies can be found in section \ref{subsec:sbi_definition}.

SNNs are a class of neural networks in which information is encoded and transmitted across neurons in a manner that more closely mimics that of the brain. Neurons in the nervous system use action potentials, rapid changes in the electrical membrane potential. These are used to trigger the neuron to transmit information to other neurons \cite{tangBridgingBiologicalArtificial2019}. Action potentials are also referred to as "spikes", a term used in the neuromorphic community. These are brief electrical events on the millisecond timescale that occur only occasionally, so neurons are typically silent. This process is dramatically more energy-efficient than the current paradigm in AI infrastructure, Artificial Neural Networks (ANNs), whose neurons compute continuously.

Although neuromorphic computing remains in its early stages, several commercially available devices have been developed, including IBM TrueNorth, Intel Loihi, and SpiNNaker2 \cite{mehonicBraininspiredComputingNeeds2022, gonzalezSpiNNaker2LargeScaleNeuromorphic2024}. These use modules of conventional Complementary Metal-Oxide-Semiconductor (CMOS) digital technologies to emulate synaptic plasticity, but are constrained by the memory-compute separation inherent to Von Neumann-style digital circuit implementations. Currently, research focuses on architectures that more closely resemble synaptic plasticity, including memristive neuromorphic systems \cite{kumarThirdorderNanocircuitElements2020,loefflerNeuromorphicLearningWorking2023,caravelliSelfOrganisingMemristiveNetworks2025}. These consist of a net of memristors, memtransistors, and resistive synapses, where neuroplasticity is encoded within the conductance and hysteresis effects of these devices \cite{zhaoSpikingArtificialNeuron2025}. However, this technology is relatively recent, and while it shows promise for the future of \textit{in-silico} neuromorphic devices, it is not yet sufficiently reliable for deployment in larger systems.

\begin{figure}[t]
    \centering
    \includegraphics[width=0.5\linewidth]{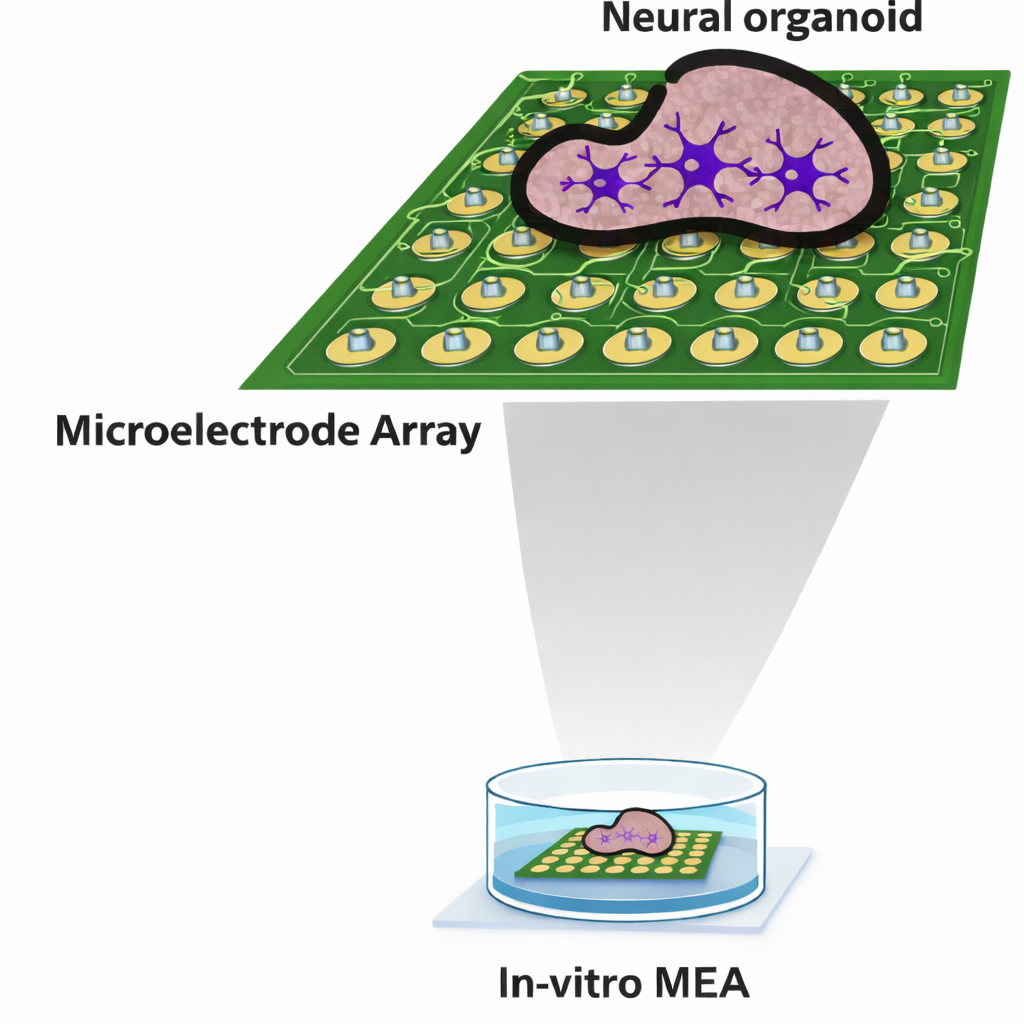}
    \caption{Representation of the in-vitro MEA, showing the neural culture (as an organoid) placed directly on top of the electrode array.}
    \label{fig:organoid_mea}
\end{figure}

\begin{figure}[t]
    \centering
    \includegraphics[width=0.7\linewidth]{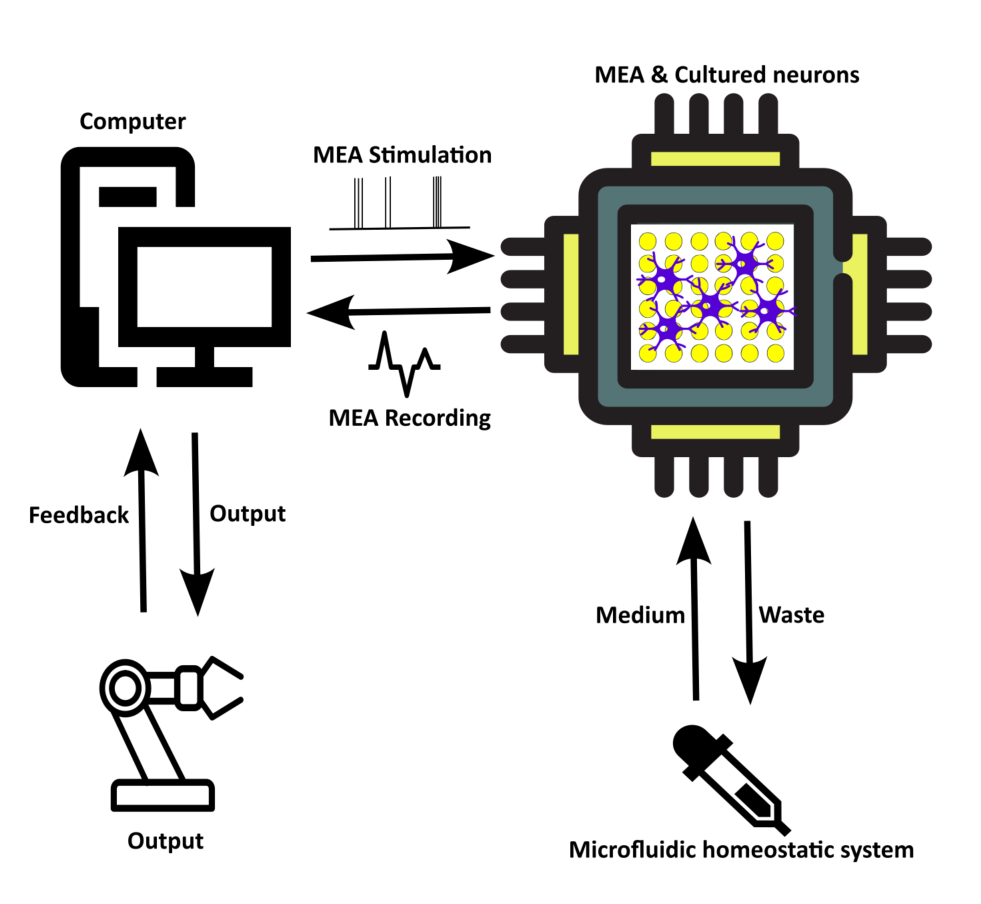}
    \caption{Schematic of an SBI setup showcasing the components of its closed-loop system. A computer sends a signal to the MEA, which then delivers an electrical stimulus to the neurons. The electrical signals from the neurons are recorded by the MEA, which then transmits them to the computer for processing. Depending on the received response, the computer sends a corresponding signal to the output to produce an appropriate action. Feedback from the action is returned to the computer, which then generates a new signal to the MEA to stimulate the neurons and improve the task's output performance. To keep the neural culture alive and functioning properly, it is maintained by a microfluidic homeostatic system that provides nutrients and removes waste. }
    \label{fig:sbi_testbed_schematic}
\end{figure}

A research area of growing prominence is the use of actual BNNs as functional components in hybrid engineered systems, a new paradigm coined SBI \cite{kaganVitroNeuronsLearn2022,kaganWhyAIProgress2025}. SBI integrates the interface of living neural cultures with hardware and software to perform task-oriented information processing in a closed loop. This has become more easily accessible with the development of induced Pluripotent Stem Cells (iPSCs) and their contact with MEAs \cite{smirnovaOrganoidIntelligenceOI2023, talaveraBrainOrganoidComputing2025}, illustrated in Fig.~\ref{fig:organoid_mea}. This permits the use of far more complex units, such as human neurons, as the computational array within this system, leveraging their intrinsic dynamics, plasticity, and adaptability, which provide remarkable efficiency. The creation of testbeds with feedback loops, the discovery of algorithms to find the most appropriate neurons to stimulate and receive information, and the spike encoding and decoding mechanisms have produced the first functional experiments of SBI \cite{tessadoriClosedloopNeuroroboticExperiments2015,kaganVitroNeuronsLearn2022}. These foundational cross-domain setups have led to more sophisticated applications, turning SBI into a popular research field. A schematic of how SBI systems work can be seen in Fig.~\ref{fig:sbi_testbed_schematic}.

In parallel to neural approaches, a broader class of biological computing paradigms has explored computation using non-neural biological substrates, including DNA computing, molecular and biochemical reaction networks, and enzyme-based logic systems \cite{BiocomputationMovingTuring2024,katzBiocomputingToolsAims2015}. These approaches exploit the inherent parallelism and information density of biochemical processes and have been extensively studied for applications such as combinatorial optimization, data storage, and molecular communication. Although these paradigms fall under the umbrella of biocomputing, they differ fundamentally from neural-based systems in substrate, dynamics, and interfacing mechanisms and are therefore outside the scope of this survey.

From a computational-theoretical perspective, SBI systems also raise the question of their relation to classical models of computation. Traditional digital computing is framed under the Turing machine abstraction \cite{maclennanNaturalComputationNonTuring2004}. Beyond-digital computing can be placed within a Beyond-Turing model, such as analog computation or real-number machines, including the Blum-Shub-Smale model \cite{blumTheoryComputationReal1988,siegelmannAnalogComputation1999}. Nevertheless, SBI systems differ fundamentally from these models: whereas these models employ algorithmic or formal abstractions, SBI systems are physical, adaptive systems \cite{jaegerFormalTheoryComputing2023}. The computational capabilities of SBI arise from learning and interaction and are not within the traditional frameworks of theoretical computation. Moreover, the intertwining of dynamics in multiple scales that occurs in BNNs, such as molecular, synaptic and global brain dynamics, increases the computing capacity over other traditional computing systems \cite{milinkovicBiologicalArtificialConsciousness2026}. Therefore, this survey does not address computational complexity or other formal descriptions; it focuses on system-level architectures and interaction schemes useful for SBI setups.

As this work focuses on computation within a system-level abstraction, all SBI testbed use cases mentioned are described primarily in this context. However, computation and system-level engineering are not the only relevant subjects that can benefit from SBI. The interdisciplinary nature of these bio-digital systems results in many different areas where investigation is beneficial. The fields of neuroscience and medical research are areas in which this technology can contribute; for example, research on neurodevelopmental or neurodegenerative diseases could be accelerated by conducting in-vitro treatment experiments rather than requiring longer and potentially dangerous testing processes \cite{yadavBrainOrganoidsTiny2021,ajongboloBrainOrganoidsAssembloids2025}.

Some preliminary testbeds had been devised for the use of SBI in systems, mainly for neurorobotics, before the launch of the first clear example of learning-like adaptation in a closed-loop environment, \textit{Dishbrain} \cite{regerConnectingBrainsRobots2000,demarseNeurallyControlledAnimat2001,novellinoConnectingNeuronsMobile2007,tessadoriModularNeuronalAssemblies2012,kaganVitroNeuronsLearn2022}. The availability of commercially available SBI platforms, such as Cortical Labs CL-1 \cite{kaganCL1PlatformTechnology2025}, as well as access to cloud-based platforms such as the Cortical Cloud or FinalSpark's \textit{Neuroplatform} \cite{jordanOpenRemotelyAccessible2024} have expanded access to this technology. However, many scientific teams lack the stringent operational requirements for conducting experiments on living neural systems, such as maintaining wet-lab facilities. 

Despite rapid advances in biological substrates and platforms, SBI remains fundamentally limited by the lack of well-defined interfacing architectures, coding schemes, and benchmarking methodologies \cite{chenOverviewVitroBiological2023}. This survey introduces a novel framework for SBI, i.e., we frame SBI systems as closed-loop bio-digital interaction architectures, where stimulation, neural dynamics, interpretation, and feedback are adapted by leveraging neuroplasticity and the substrate's learning capabilities. This departs from Shannon's conception of stationary, memory-limited, and symbol-centered channels. By doing so, we provide a system-level protocol that enables standardized analysis, comparison, and benchmarking of SBI platforms.

Building on this abstraction, this survey reviews existing SBI frameworks through the interaction architecture, signal encoding, modulation, and decoding, and the performance metrics used to benchmark them, which vary across experiments. We then present open questions and future challenges that face the application and expansion of SBI platforms. Finally, we offer a perspective on future directions for establishing SBI as the most valuable component of next-generation hybrid systems. 

\subsection{Definition and Scope of Synthetic Biological Intelligence}
\label{subsec:sbi_definition}

So far, there have been conflicting definitions regarding the correct terminology for what we consider SBI. In \cite{kaganTechnologyOpportunitiesChallenges2023} and \cite{mansanoriMoralDimensionsSynthetic2024}, the authors define SBI as the intentional synthesis of a combination of biological and silicon substrates in-vitro for the purpose of goal-directed or otherwise intelligent behavior. The work in \cite{sellarCyberneticFrameworkSynthetic2025} defines SBI as an engineered system that leverages biocomputation to produce cognitive functions. In \cite{tanveerStartingSyntheticBiological2025}, SBI is defined as the assessment of the functional information processing capabilities of in-vitro BNNs via electrophysiological activity to elicit emergent intelligent behaviors. One of the most exhaustive works on terminology in the biocomputing field, in \cite{patelComputationalPerspectiveNeuroAI2025}, describes SBI as the construction of hybrid systems that incorporate living neural cells as computational substrates. 

In this survey, we use the term SBI to denote systems in which living BNNs are interfaced with hardware and software to perform task-oriented information processing in a closed loop. This definition emphasizes the use of biological neural substrates as the primary components of engineered systems, rather than the biology of the neurons themselves as the focus of investigation \cite{kaganVitroNeuronsLearn2022}. 

\begin{figure*}[t]
\centering
\includegraphics[width=0.7\linewidth]{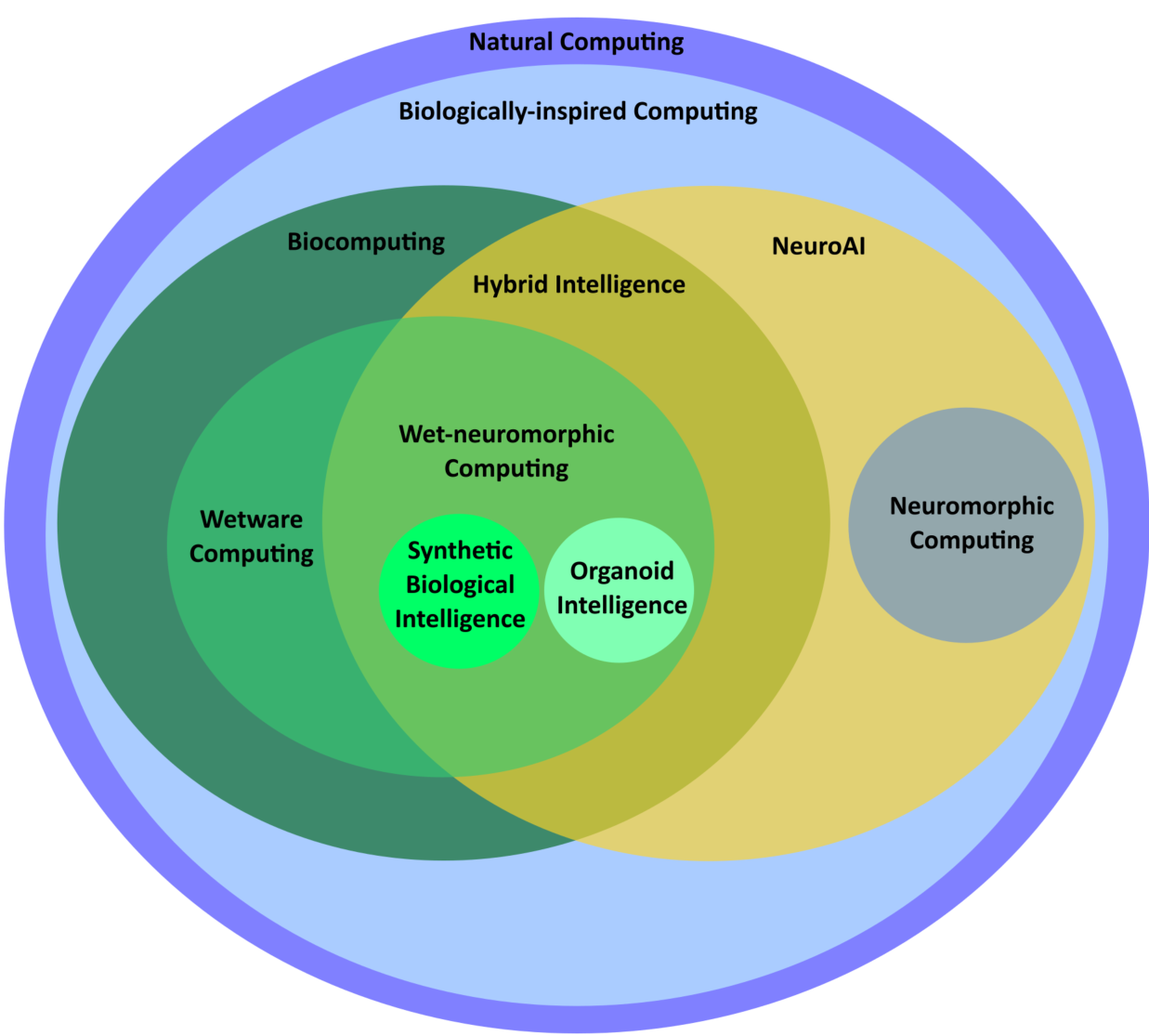}
    \caption{Hierarchical relationship between the terminology in biologically-inspired computing.}
    \label{fig:bio_computing_hierarchy}
\end{figure*}

Other terms for related research include organoid intelligence (OI) \cite{smirnovaOrganoidIntelligenceOI2023,ballavOrganoidIntelligenceBridging2024,liOrganoidComputingLeveraging2024}, \textit{NeuroAI} \cite{patelComputationalPerspectiveNeuroAI2025}, \textit{wetware computing} \cite{liAdvancedBrainonaChipWetware2025}, \textit{hybrid intelligence} \cite{shaoOpportunitiesChallengesBrainonaChip2025}, \textit{biocomputing}, \textit{neuromorphic computing}, among others. We provide an overview of all related terms for clarification in Table~\ref{tab:neuro_terms} and a visual representation of how they relate to each other in Fig. \ref{fig:bio_computing_hierarchy}. These terms often emerge from diverse research teams that focus on different applications and configurations of biological computation. 

\begin{table*}[t]
\caption{Terminology and paradigms in the biologically-inspired computing community.}
\label{tab:neuro_terms}
\centering
\renewcommand{\arraystretch}{1.2}
\begin{tabular}{m{2.8cm}m{5cm}m{3.8cm}m{2.2cm}m{1.8cm}}
\hline
 \textbf{Terminology} 
& \textbf{Description} 
& \textbf{Configuration} 
& \textbf{Scale}
& \textbf{References} \\
\hline

Natural computation
& Umbrella term for all previous terms. Computation including or inspired by nature.
& Physical, chemical, or biological natural processes at multiple scales.
& All scales
& \cite{maclennanNaturalComputationNonTuring2004} \\
\hline

Biologically-inspired computing
& Insights from all biological systems inspire the design of intelligent schemes, both artificial and biological.
& Biological principles are implemented in digital or analog systems.
& System
& \cite{karBioInspiredComputing2016} \\
\hline

Biocomputing
& Biological cells or organisms are used to perform computational tasks with their natural molecular functions.
& Molecular, biochemical, or cellular substrates (such as cells, DNA, proteins).
& Molecular / Cellular/ Network
& \cite{katzEnzymeBasedLogicSystems2010,katzBiocomputingToolsAims2015} \\
\hline

Wetware Computing
& Living biological material is used as a substrate for information processing. Includes SBI, as well as intracellular processes and open-loop systems such as reservoir computing.
& Living biological substrates with external stimulation interfaces.
& Cellular / Network
& \cite{liAdvancedBrainonaChipWetware2025} \\
\hline

Hybrid Intelligence
&  Systems where part of the AI functions are executed in biological cells, neuro-inspired but not necessarily with neurons, and interfaced with conventional computing devices.
& Bio-digital systems combining biological and artificial computational modules.
& System
& \cite{shaoOpportunitiesChallengesBrainonaChip2025, pereraWetNeuromorphicComputingNew2025} \\
\hline

Wet-Neuromorphic Computing
& Computation is performed by living biological systems that seek to mimic neural computation.
& Living neurons or cell cultures organized to emulate neural architectures.
& Cellular / Network
& \cite{pereraWetNeuromorphicComputingNew2025} \\
\hline

Synthetic Biological Intelligence
& Closed-loop systems where living BNNs are interfaced with digital components to perform task-oriented information processing.
& In-vitro living neural networks with closed-loop digital control.
& Network
& \cite{barrosEditorialIntersectionBiological2025,hofmannControllingCyberPhysicalSystem2025,kaganHarnessingIntelligenceBrain2025,patelComputationalPerspectiveNeuroAI2025} \\
\hline

Organoid Intelligence
& Computational paradigms using organoids as biological substrates. The focus is on biological structure and developmental investigation rather than on task performance.
& Three-dimensional organoids with electrophysiological or chemical interfacing.
& Network
& \cite{smirnovaOrganoidIntelligenceOI2023,ballavOrganoidIntelligenceBridging2024,liOrganoidComputingLeveraging2024} \\
\hline

Neuromorphic Computing
& Silicon-based computational systems inspired by neural processes.
& In-silico SNNs in non von-Neumann microprocessor architectures.
& Network
& \cite{ranaNeuralVsNeuromorphic2025} \\
\hline

NeuroAI
& Neuroscience insights inspire the design of intelligent systems, both artificial and biological.
& Neuroscience-guided algorithms, hardware and software architectures, and learning principles.
& System
& \cite{patelComputationalPerspectiveNeuroAI2025} \\
\hline

Brain-Machine Interface (BMI)
& In-vivo neural interfaces are used to improve cognitive and sensorimotor function.
& Implanted or non-invasive interfaces with living nervous systems in-vivo.
& System
& \cite{shaoOpportunitiesChallengesBrainonaChip2025} \\
\hline

\end{tabular}
\end{table*}


\begin{figure*}[t]
\centering
\includegraphics[width=0.8\linewidth]{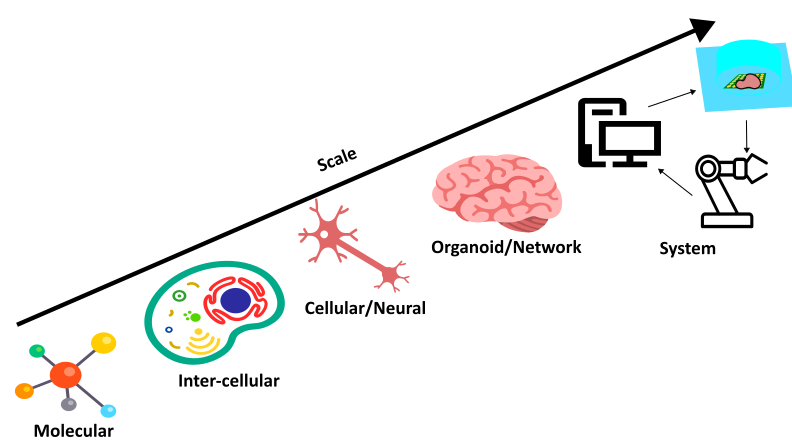}
    \caption{\centering Schematic of scale representation in biologically-inspired computing.}
    \label{fig:bio_computing_scales}
\end{figure*}

The terminology used throughout the biologically inspired computing literature varies not only in substrate but also in the spatial and organizational scales at which computation emerges. At smaller spatial scales, computation is realized through molecular or cellular interactions, including neural or non-neural cells whose biophysical properties perform information processing. Meanwhile, at larger scales, computation arises from coordinated network dynamics within neural populations and from the closed-loop interactions of whole systems, such as whole SBI testbeds with embodied outputs or bio-digital interfaces. This level separation can be visualized in Fig. \ref{fig:bio_computing_scales}, where the different scales at which biologically-inspired computing occurs are contrasted. In this work, we use the notion of scale to distinguish between computational paradigms operating at different levels of biological organization, and to clarify the system-level nature of SBI.

Within the scope of biologically inspired computing, terms such as OI and wetware computing can be viewed as domain-specific instances or conceptual variants of SBI, differing primarily in the biological substrate, scale, or application focus rather than in underlying system principles \cite{barrosEditorialIntersectionBiological2025}. Neuromorphic computing is inspired by neural processes, as these systems use spike-based representations in their hardware, though they are silicon-based rather than biological. Biocomputing is an umbrella term for all computation involving biological processes, including SBI, DNA, and molecular computing, among others. Wetware computing, which uses living biological material as a substrate, encompasses SBI, reservoir computing, and intracellular processes. When we focus on organoids as the biological systems whose computational properties are leveraged, we refer to them as OI. We only consider it SBI when these are embedded in closed-loop interfaces for information processing. 

SBI also shares several similarities with \textit{Brain-Machine Interfaces} (BMIs), and many SBI techniques are derived from these developments. Nevertheless, BMIs primarily refer to \textit{in-vivo} schemes designed to restore cognitive and sensorimotor function \cite{shaoOpportunitiesChallengesBrainonaChip2025}.

\subsection{Historical Development of SBI}
\label{subsec:sbi_history}

 In the early 2000s, Shahaf and Marom first demonstrated feedback-driven learning in in-vitro cortical networks \cite{shahafLearningNetworksCortical2001}. Soon after, DeMarse \emph{et al.} used these networks in closed-loop systems with an external task embodiment, thereby establishing the first SBI systems \cite{demarseNeurallyControlledAnimat2001}. 

For the following two decades, the focus was shifted to neurorobotics, with research groups creating varied embodied testbeds, both physical \cite{novellinoConnectingNeuronsMobile2007,warwickControllingMobileRobot2010,shultza.lees.sheat.b.&yancoh.a.ControlRobotArm2014,itoChangesNetworkActivity2018} and virtual \cite{demarseAdaptiveFlightControl2005,ruaroNeurocomputerImageProcessing2005,bakkumSpatiotemporalElectricalStimuli2008}. These systems were instrumental in establishing closed-loop learning paradigms between biological substrates and artificial environments, laying essential groundwork for future SBI architectures.

In the late 2010s, animal and human iPSC-derived neurons enabled the generation of neurons in-vitro. These were primarily intended for neurodevelopmental studies and disease modeling. Subsequent improvements in differentiation protocols have enabled the development of more mature and functional neural networks \cite{forroModularMicrostructureDesign2018,andersenGenerationFunctionalHuman2020}. Combined with advances in MEAs, which increased electrode density, spatial resolution, and stimulation fidelity  \cite{massobrioVitroStudiesNeuronal2015,napoliInvestigatingBrainFunctional2015,zaniniInvestigatingReliabilityEvoked2023}, these developments significantly improved the controllability, interpretability, and scalability of neural cultures. 

\subsection{Related Work}
\label{subsec:related_work}

This subsection briefly describes and compares existing related SBI review papers and outlines the contribution of this survey. Thus, Table ~\ref{tab:comparison_reviews} compares related work in various categories.

The works in \cite{bisioClosedLoopSystemsVitro2019} and \cite{georgePlasticityAdaptationNeuromorphic2020} do not explicitly mention the concept of SBI (nor one of the related terms); however, a summary of existing BNN testbeds in closed-loop configurations for task learning is given. Also, not considering a term for SBI, in \cite{chenOverviewVitroBiological2023}, the authors focus on BNN testbeds for robotic intelligence applications, mentioning their encoding and decoding techniques.

The first word mentioning a concept that encompasses these systems, \textit{Organoid Intelligence} (OI), is made by Smirnova \emph{et al.} in \cite{smirnovaOrganoidIntelligenceOI2023}, in which existing setups are listed.
The authors provide a definition of OI, the advances that enable its feasibility, and a blueprint for its future application. In \cite{ballavOrganoidIntelligenceBridging2024}, the authors also provide a review of OI in the context of the history of biocomputing and its applications for medicine. 
\cite{liOrganoidComputingLeveraging2024} and \cite{wadanOrganoidIntelligenceBiocomputing2025} review \textit{organoid computing} in the scope of neuromorphic computation, and center more on the computing side rather than the biological side. However, these works do not address the interaction system, which is examined in subsequent reviews.

The work in \cite{patelComputationalPerspectiveNeuroAI2025} provides a detailed definition of SBI, how the different configurations relate to learning strategies, and the architectures that compose them. The authors in \cite{talaveraBrainOrganoidComputing2025} focus on highlighting the characteristics of SBI systems and how these can be leveraged for computing applications.

The review presented in \cite{huaMicroelectrodeArraysCultured2025} surveys BNN testbeds in MEAs, explaining interaction system architectures, encoding and decoding techniques, and the testbeds on which they were applied. However, they do not explicitly mention a concept that encompasses these neurocomputing systems.

One of the most complete SBI reviews is provided in \cite{liAdvancedBrainonaChipWetware2025}. This review covers the interfacing aspects of wetware computing, provides examples of existing schemes, and benchmarks their performance against ANNs. However, it does not provide a definition of SBI. The authors in \cite{shaoOpportunitiesChallengesBrainonaChip2025} also provide a review of brain-on-a-chip systems, focusing on interaction architectures and interfaces.

In contrast to existing surveys, which focus primarily on biological substrates, neuromorphic computation, or isolated experimental platforms, our work frames SBI in the scope of a system-level interaction paradigm. This includes encoding and modulation schemes, decoding and detection techniques, adaptive substrate characteristics, types of noise, and performance metrics. Furthermore, this survey emphasizes system-level integration and reproducibility, highlighting the need for standardized interfaces, benchmarking methodologies, and evaluation protocols. A platform-integrated case study is included to illustrate an end-to-end SBI interaction pipeline, without restricting the discussion to a single implementation. As such, this work aims to bridge the gap between SBI experimentation and systems engineering.

\begin{table*}[t]

\caption{Comparison of this work with existing surveys and reviews in SBI, listed in chronological order. Symbols indicate the evaluation of the categories: \cmark (fully addressed), \pmark (partially addressed), and \xmark (not addressed).}
\label{tab:comparison_reviews}
\centering
\renewcommand{\arraystretch}{1.2}
\setlength{\tabcolsep}{2pt}

\begin{tabular}{lccccccccccccccc}
\hline
\textbf{Category / Reference} & \textbf{\cite{bisioClosedLoopSystemsVitro2019}} & \textbf{\cite{georgePlasticityAdaptationNeuromorphic2020}} & \textbf{\cite{smirnovaOrganoidIntelligenceOI2023}} & \textbf{\cite{chenOverviewVitroBiological2023}} & \textbf{\cite{ballavOrganoidIntelligenceBridging2024}} & \textbf{\cite{liOrganoidComputingLeveraging2024}} & \textbf{\cite{wadanOrganoidIntelligenceBiocomputing2025}} &\textbf{\cite{patelComputationalPerspectiveNeuroAI2025}} &  \textbf{\cite{talaveraBrainOrganoidComputing2025}}  & \textbf{\cite{huaMicroelectrodeArraysCultured2025}} & \textbf{\cite{yaronDissociatedNeuronalCultures2025}} &\textbf{\cite{liAdvancedBrainonaChipWetware2025}} & 
\textbf{\cite{shaoOpportunitiesChallengesBrainonaChip2025}} & \textbf{\cite{sellarCyberneticFrameworkSynthetic2025}}  & \textbf{This Work} \\
\hline
Year                                 & 2019      & 2020      & 2023  & 2023      & 2024      & 2024     & 2025      & 2025    & 2025    & 2025  & 2025     & 2025      & 2025  & 2025    & 2026 \\
SBI definition and scope             & \xmark    & \xmark    & \cmark    & \xmark    & \cmark   & \cmark   & \cmark    & \cmark & \xmark    & \xmark    & \pmark  & \pmark & \cmark     & \cmark & \cmark \\
SBI platforms and substrates         & \xmark    & \pmark    & \cmark  & \pmark   & \cmark    & \cmark   & \cmark   & \cmark    & \pmark & \pmark    & \pmark    & \cmark  & \cmark & \pmark    & \cmark \\
Configuration        & \cmark    & \xmark    & \pmark  & \pmark   & \xmark    & \xmark   & \xmark   & \pmark    & \pmark & \cmark    & \pmark    & \cmark  & \cmark & \pmark    & \cmark \\
Encoding and modulation schemes      & \xmark    & \pmark    & \xmark & \pmark   & \xmark    & \xmark   & \xmark   & \xmark    & \xmark & \pmark    & \xmark    & \cmark  & \cmark & \xmark   & \cmark \\
Decoding and detection techniques    & \cmark    & \xmark    & \xmark  & \pmark   & \xmark    & \xmark   & \xmark   & \pmark    & \xmark & \pmark    & \xmark    & \cmark  & \pmark & \xmark    & \cmark\\
Benchmarking metrics and evaluation  & \xmark    & \xmark    & \xmark & \xmark    & \xmark    & \xmark   & \xmark   & \xmark    & \xmark & \xmark    & \xmark    & \cmark  & \xmark & \xmark    & \cmark \\
Testbeds       & \cmark    & \cmark    & \pmark    & \pmark    & \cmark & \cmark   & \xmark   & \pmark    & \xmark & \cmark    & \cmark    & \pmark  & \cmark & \pmark    & \cmark \\
Reproducibility and standardization  & \xmark    & \xmark    & \xmark  & \xmark   & \xmark    & \xmark   & \xmark   & \xmark    & \xmark & \xmark    & \pmark    & \pmark  & \xmark &  \xmark  & \cmark \\
\hline
\end{tabular}
\end{table*}

\subsection{Paper Organization}\label{subsec:paper_organization}

The remainder of this survey is as follows: Section~\ref{sec:sbi_survey} provides an overview of the neural cultures used for SBI, the early stage testbeds and the current state-of-the-art.
Section~\ref{sec:abnia} introduces the interfacing framework proposed for SBI, which is called the Adaptive Bio-Neural Interaction Architecture (ABNIA). 
Section~\ref{sec:coding_schemes} comprises the encoding and modulation methods. Section~\ref{sec:channel_characteristics} provides an overview of the emergent properties of neural substrates and the considerations for including them in systems. In Section~\ref{sec:decoding_detection}, we present the decoding schemes, the feedback constraints, and the metrics that can be used for benchmarking. Potential challenges, future applications, and research directions are discussed in Section~\ref{sec:future_work}. Finally, Section~\ref{sec:conclusion} summarizes the implications of this SBI framework and provides a conclusion. 

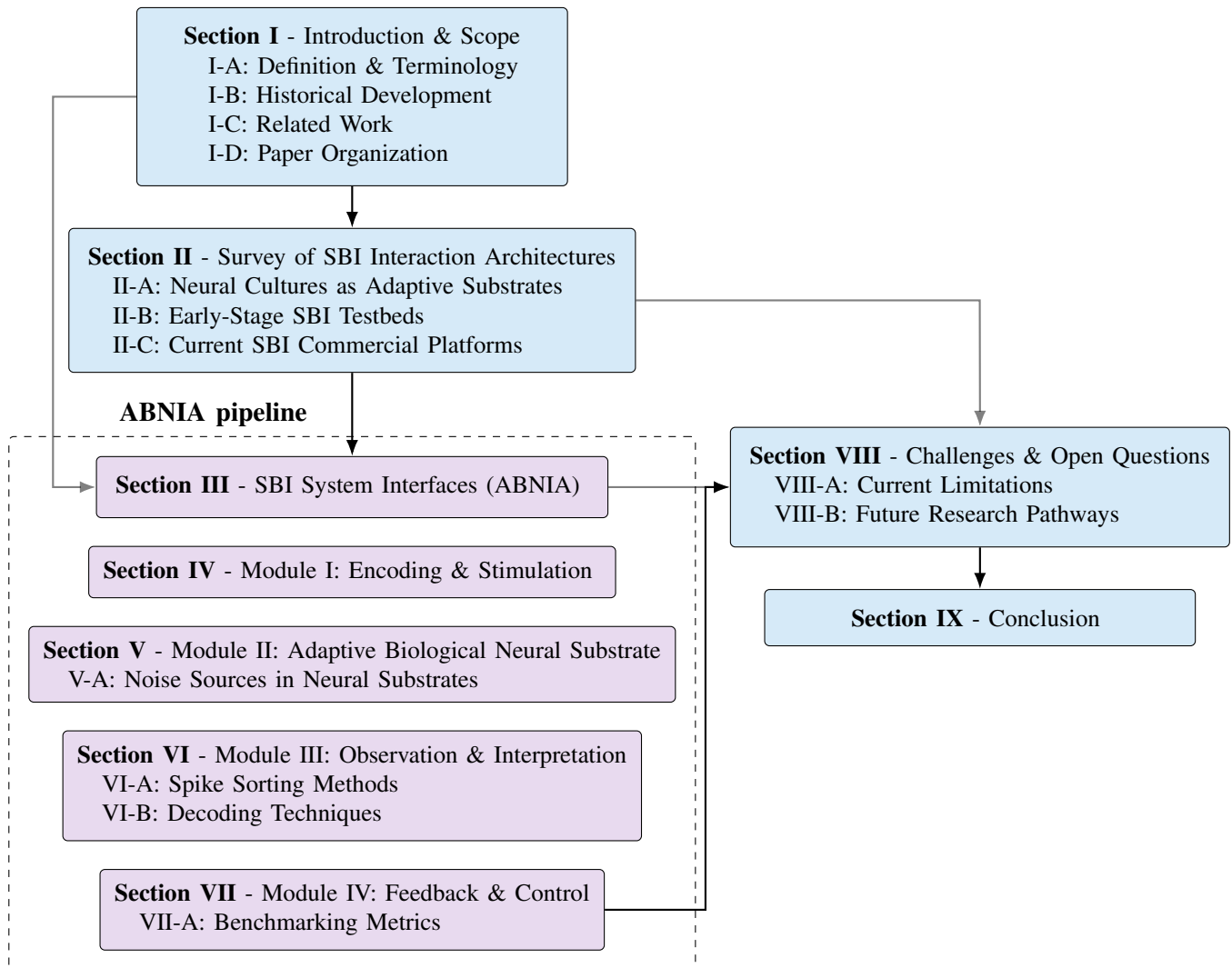
\begin{figure}[t]
\centering
\resizebox{\linewidth}{!}{
\begin{tikzpicture}[
  font=\small,
  box/.style={
    draw, rounded corners=2pt, align=left,
    minimum width=6.2cm, inner sep=6pt
  },
  bigbox/.style={
    draw, rounded corners=2pt, align=left,
    minimum width=6.2cm, inner sep=8pt
  },
  arrow/.style={-Latex, thick},
  dashedgroup/.style={draw, dashed, rounded corners=2pt, inner sep=8pt},
  title/.style={font=\bfseries}
]

\node[bigbox, fill=introcolor] (intro) {
  {\textbf{Section I} - Introduction \& Scope}\\
  \quad I-A: Definition \& Terminology\\
  \quad I-B: Historical Development\\
  \quad I-C: Related Work\\
  \quad I-D: Paper Organization
};

\node[bigbox, fill=introcolor, below=6mm of intro] (survey) {
  {\textbf{Section II} - Survey of SBI Interaction Architectures}\\
  \quad II-A: Neural Cultures as Adaptive Substrates\\
  \quad II-B: Early-Stage SBI Testbeds\\
  \quad II-C: Current SBI Commercial Platforms
};

\node[bigbox, fill=archcolor, below=12mm of survey] (abnia) {
  {\textbf{Section III} - SBI System Interfaces (ABNIA)}
};

\node[box, fill=archcolor, below=4mm of abnia] (enc) {
  {\textbf{Section IV} - Module I: Encoding \& Stimulation}
};

\node[box, fill=archcolor, below=4mm of enc] (sub) {
  {\textbf{Section V} - Module II: Adaptive Biological Neural Substrate}\\
  \quad V-A: Noise Sources in Neural Substrates
};

\node[box, fill=archcolor, below=4mm of sub] (obs) {
  {\textbf{Section VI} - Module III: Observation \& Interpretation}\\
  \quad VI-A: Spike Sorting Methods\\
  \quad VI-B: Decoding Techniques
};

\node[box, fill=archcolor, below=4mm of obs] (fbk) {
  {\textbf{Section VII} - Module IV: Feedback \& Control}\\
  \quad VII-A: Benchmarking Metrics
};

\node[dashedgroup, fit=(abnia)(enc)(sub)(obs)(fbk), label={[title, xshift=-20mm]above:ABNIA pipeline}] (group) {};


\node[bigbox, fill=introcolor, right=17.5mm of abnia] (challenges) {
  {\textbf{Section VIII} - Challenges \& Open Questions}\\
  \quad VIII-A: Current Limitations\\
  \quad VIII-B: Future Research Pathways
};

\node[bigbox, fill=introcolor, below=6mm of challenges] (conc) {
  {\textbf{Section IX} - Conclusion}
};

\draw[arrow] (intro) -- (survey);
\draw[arrow] (survey) -- (abnia);
\draw[arrow] (fbk.east) --  ++(14.5mm,0) |- (challenges.west);         
\draw[arrow] (challenges) -- (conc);

\draw[arrow, opacity=0.5] (intro.west) -- ++(-12mm,0) |- (abnia.west);
\draw[arrow, opacity=0.5] (survey.east) -- ++(49.5mm,0) -- (challenges.north);
\draw[arrow, opacity=0.5] (abnia.east) -- ++(14mm,0) |- (challenges.west);

\end{tikzpicture}
}
\caption{Structure of the survey and recommended reading paths. The dashed block highlights the ABNIA interaction pipeline used as the paper’s system-level abstraction.}
\label{fig:paper_structure}
\end{figure}

To provide readers with a roadmap for reading the survey, including suggested reading paths, Fig.~\ref{fig:paper_structure} presents an overview of the structure. We also provide a table of abbreviations for ease of reading in Tab. \ref{tab:abbreviations}. 


\begin{table}[t]
\centering
\caption{Alphabetical List of Abbreviations Used Throughout This Survey}
\label{tab:abbreviations}
\renewcommand{\arraystretch}{1.1}
\begin{tabular}{ll}
\hline
\textbf{Abbreviation} & \textbf{Definition} \\
\hline
\textbf{ABNIA} & Adaptive Bio-Neural Interaction Architecture \\
\textbf{AFR} & Average Firing Rate \\
\textbf{AI} & Artificial Intelligence \\
\textbf{ANN} & Artificial Neural Network \\
\textbf{BMI} & Brain-Machine Interface \\
\textbf{BNN} & Biological Neural Network \\
\textbf{CMOS} & Complementary Metal-Oxide Semiconductor \\
\textbf{DNA} & Deoxyribonucleic Acid \\
\textbf{EM} & Expectation-Maximization \\
\textbf{IIR} & Infinite Impulse Response \\
\textbf{iPSC} & Induced Pluripotent Stem Cell \\
\textbf{ISI} & Inter-Spike Interval \\
\textbf{KLD} & Kullback--Leibler Divergence \\
\textbf{MEA} & Microelectrode Array \\
\textbf{MI} & Mutual Information \\
\textbf{OI} & Organoid Intelligence \\
\textbf{OOK} & On-Off Keying \\
\textbf{PCA} & Principal Component Analysis \\
\textbf{PSTH} & Peri-Stimulus Time Histogram \\
\textbf{SBI} & Synthetic Biological Intelligence \\
\textbf{SNN} & Spiking Neural Network \\
\textbf{STDP} & Spike-Timing-Dependent Plasticity \\
\textbf{SVM} & Support Vector Machine \\
\textbf{TTFS} & Time-to-First-Spike \\
\hline
\end{tabular}
\end{table}

\section{Survey of SBI Architectures}
\label{sec:sbi_survey}

The emergence of SBI has been driven by advances in multiple enabling technologies, most notably the development of \textit{in-vitro} neural cultures and MEA systems. Together, these advances have enabled the controlled stimulation, recording, and interfacing of living BNNs, forming the foundation for contemporary SBI platforms.

Most SBI studies have historically focused on task-driven learning in closed-loop systems, often using robotic embodiments—both virtual and physical—as intuitive performance benchmarks. More recently, this scope has expanded to include cloud-accessible and commercial SBI platforms, which enable systematic investigation of these interfaces beyond embodied control tasks.

\subsection{Neural Cultures as Adaptive Substrates}
\label{subsec:neural_cultures}

The first dissociated neural culture was that of 400 neurons of a chick embryo in 1974, for which the electrical properties of cells were measured under certain chemicals and electrical stimuli \cite{chalazonitisElectrophysiologicalCharacteristicsChick1974}. However, these early cultures were primarily observational tools, not designed for information processing.

Following these pioneering works, cell culture techniques made continuous advances, which along with improvements in electrophysiological interfacing, allowed the long-term maintenance of neural networks in-vitro \cite{pineRecordingActionPotentials1980,potterNewApproachNeural2001,halesHowCultureRecord2010}. The main focus of the subsequent decades was to enable stable recording of neural activity, albeit limited until the introduction of MEAs, which greatly increased the precision of the obtained data and enabled stimulation and closed-loop interaction within the culture.

Access to different types of animal neurons was restricted because it required brain dissection. Therefore, rodents have been preferred for the extraction of neural substrates, given their wide availability in research \cite{shahafLearningNetworksCortical2001,novellinoBehaviorsElectricallyStimulated2003,warwickControllingMobileRobot2010}. The development of iPSCs, in particular that of human induced pluripotent stem cells (hiPSC) \cite{takahashiInductionPluripotentStem2007}, led to a boost in neuroscientific research. This technology enables modeling diverse disorders and the use of human neurons, which were previously limited by the invasiveness of extraction.

Most recently, neural cultures used in SBI systems have been obtained through different methods. SBI studies have used both animal-derived neurons, most commonly rodent cells, and human-derived neural cultures. While direct comparisons remain limited, several studies report differences in learning dynamics and adaptation capabilities across cell types, suggesting that substrate biology plays a significant role in shaping SBI channel properties \cite{kaganVitroNeuronsLearn2022}. In particular, neurons of the human cerebral cortex have been widely used as a substrate, as they are highly evolved regions of the brain \cite{qianBrainOrganoidsAdvances2019}. The differentiation technique used to generate iPSCs also affects the properties of the neurons themselves, and therefore the entire culture \cite{napoliComparativeAnalysisHuman2016}. 

The most widely used SBI substrates are dissociated two-dimensional cultures, but research is currently investigating more sophisticated three-dimensional neural substrates \cite{janzenCorticalNeuronsForm2020}. The latter considers brain organoids \cite{qianBrainOrganoidsAdvances2019}, which offer a more complex architecture and can more reliably approximate brain structure \cite{bakkumRemovingAIEmbodied2004,muzziHumanDerivedCorticalNeurospheroids2023}. Three-dimensional MEAs and polymer-based neural scaffolds are currently developing technologies that aim to improve stimuli and recording capabilities for more complex cultural structures \cite{obienRevealingNeuronalFunction2015,hagiwaraFabricationTraining3D2023,abushihadaHighlyCustomizable3D2024,upadhyayLivingNeurosheetsEngineering2025,jungFlexible3DKirigami2025}.

Unlike artificial computational substrates, neural cultures are living systems that require continuous maintenance. This includes temperature control, nutrient-rich culture media, gas exchange, and waste management \cite{killianDeviceLongTermPerfusion2016,kaganHarnessingIntelligenceBrain2025}. Also, cultures evolve over time, undergoing developmental phases in which synapses form and electrical activity is nascent, and eventually age and degrade. SBI substrates are mortal \cite{talaveraBrainOrganoidComputing2025}, currently having a lifespan of a couple of months. This introduces practical constraints, especially in long-term experiments and reproducibility.

Neural cultures exhibit several properties that distinguish them from conventional computational substrates. These include spontaneous activity, long-term and short-term synaptic plasticity, strong temporal memory, nonlinearity, and population-level dynamics such as synchronization and bursting. These properties enable adaptive, and learning-driven behavior but also introduce variability, drift, and sensitivity to initial conditions, complicating their use as stable system substrates.


\subsection{Early-Stage SBI Testbeds}
\label{subsec:early_sbi_testbeds}

The first implementation of a system that integrated an in-vitro neural culture with an engineered system was reported in \cite{regerConnectingBrainsRobots2000} in 2000, in which reticular neurons from a lamprey were connected to a mobile robot. For this project, the focus was on computational properties and neural preparation; nevertheless, it is considered the first prototype of an SBI system interface.

DeMarse \emph{et al.} produced the first neurohybrid system controlling a virtual simulation, called \textit{Animat} \cite{demarseNeurallyControlledAnimat2001}. Subsequently, they conducted a flight-control simulation to demonstrate the capabilities of neural cultures \cite{demarseAdaptiveFlightControl2005}. Bakkum \emph{et al.} followed with an artistic educational project, \textit{MEART}, in which the neural spikes of rat cortical neurons in a MEA were used to produce movement of ink markers on paper in different art exhibitions \cite{bakkumRemovingAIEmbodied2004,bakkumMEARTSemilivingArtist2007}. These drawings were filmed in real time to provide electrical stimulus feedback to the neurons.

After these pioneering testbeds, the focus shifted to building systems with physical robots to consolidate protocols for signal encoding and decoding, thereby improving system performance and efficiency. The laboratory of Martinoia and Chiappalone carried out several projects in this respect \cite{martinoiaEmbodiedVitroElectrophysiology2004,cozziCodingDecodingInformation2005,novellinoConnectingNeuronsMobile2007,tessadoriModularNeuronalAssemblies2012,tessadoriClosedloopNeuroroboticExperiments2015,buccelliClosedloopElectrophysiologyPresent2018,itoChangesNetworkActivity2018,zaniniInvestigatingReliabilityEvoked2023}, as did other research teams \cite{bakkumSpatiotemporalElectricalStimuli2008,kudohVitroidRobotSystem2011,shultza.lees.sheat.b.&yancoh.a.ControlRobotArm2014,liApplicationHierarchicalDissociated2015,liNovelRobotSystem2016,aaserMakingCyborgClosedloop2017}.

Several other testbeds have also been developed to address non-robotic tasks. The first was developed by Ruaro \emph{et al.} in 2005 to perform image processing \cite{ruaroNeurocomputerImageProcessing2005}. The authors in \cite{isomuraCulturedCorticalNeurons2015} performed blind-source separation experiments to test the free energy principle and the clustering capabilities of cultured neurons. Cai \emph{et al.} have more recently done experiments with speech processing and predicting non-linear equations
\cite{caiBrainOrganoidComputing2023,caiBrainOrganoidReservoir2023}.

\subsection{Current SBI Commercial Platforms}
\label{subsec:sbi_platforms}

The first commercial platform for in-vitro neural cultures, which could then be set in a closed-loop system was Neurorighter \cite{rolstonNeuroRighterClosedLoopMultielectrode2009}. Although this platform was primarily intended for neuroscience and biomedical applications, research teams also performed hardware control and robotic embodiment \cite{newmanClosedLoopMultichannelExperimentation2013}.

The developments described in Sections \ref{subsec:sbi_history} and \ref{subsec:early_sbi_testbeds} led to Kagan \emph{et al.}'s groundbreaking work, the 2022 paper in which they taught DishBrain, a human neural culture, how to play the computer game 'Pong' \cite{kaganVitroNeuronsLearn2022}. They demonstrated that external feedback modulation in a real-time closed-loop interaction can be used for creating reliable and stable SBI systems. Thus, demonstrating that these neural cultures could be used effectively as computational systems, leveraging the theoretical benefits of BNNs \cite{jeonDistinctivePropertiesBiological2023,robbinsGoalDirectedLearningCortical2024,watmuffDrugTreatmentAlters2025,kaganHarnessingIntelligenceBrain2025}.

Further experiments were conducted with Dishbrain \cite{habibollahiCriticalDynamicsArise2023,khajehnejadComplexNetworkDynamics2023,khajehnejadBiologicalNeuronsCompete2024}, to test its adaptability and measure its properties when affecting key patterns of BNN functioning, such as criticality and feedback learning patterns.

The company behind Dishbrain, Cortical Labs, further developed the testbed into the first commercially available SBI platform, the CL-1 \cite{kaganCL1PlatformTechnology2025}. This is to our knowledge the first standardized and scalable SBI system that enables a programmable interaction and computation substrate with living neural cultures \cite{hofmannControllingCyberPhysicalSystem2025,hoganCLAPIRealTime2026}.

 The CL-1 is also accessible via the cloud, through the Cortical Cloud service. Another company that offers SBI platforms with the cloud is FinalSpark with its NeuroPlatform \cite{jordanOpenRemotelyAccessible2024,liuEncodingTactileStimuli2025}. This enables infrastructure for teams worldwide without the need for wet-lab equipment, which is required to support neuronal life.

A comparison of platforms with the capacity of the human brain as a benchmark is provided in Table~\ref{tab:cl1_neuroplatform_brain}. Although SBI systems differ fundamentally from in-vivo neural systems, given their complexity, embodiment, size, and purpose, the human brain represents the only known large-scale BNN capable of robust, lifelong, and adaptive learning and information processing. Therefore, we use it as a reference point to contextualize how complex and efficient current SBI platforms are and how their orders of magnitude differ, rather than using it as a benchmark or direct target.

There are also research teams that are exploring learning procedures in neural organoids and developing reproducible and standardized SBI platforms. Tal Sharf's \textit{Braingeneers} team at UC Santa Cruz investigates feedback-driven learning and goal-directed behavior in cortical organoids, focusing on adaptive network dynamics arising from external stimuli in a closed-loop system \cite{robbinsGoalDirectedLearningCortical2024, voitiukFeedbackdrivenBrainOrganoid2025}. The University of Illinois Urbana-Champaign's \textit{Mind in-vitro} initiative is developing scalable and reproducible neural substrates, particularly \textit{neurosheets} and versatile platforms for applying stimuli and obtaining relevant signals. \cite{zhangMindVitroPlatforms2024,upadhyayLivingNeurosheetsEngineering2025}. Lena Smirnova's lab at Johns Hopkins University has contributed greatly to the field of OI, by focusing on the neurodevelopmental and neuroscientific aspects of biocomputing and organoid development, which in turn advance SBI research  \cite{smirnovaBiocomputingOrganoidIntelligence2024,alameldinHumanNeuralOrganoid2025,acha3DNeuromodulationNeural,barrosEditorialIntersectionBiological2025,smirnovaBiocomputingOrganoidIntelligence2024}.

In addition, there are two startups that have emerged from universities. The first is Neurorium, originating from the Freie Universität Berlin. It focuses on developing BNNs as biological processors to find energy-efficient alternatives to in-silico computing.  \cite{venkinaBioprozessorenKoennenMiniGehirne2026}. Neurodelphus, located in Philadelphia, is supported by Thomas Jefferson University and focuses on BMI for neurological trauma and disease, and to build novel computational systems with closed-loop stimulation and interactions between heterogeneous neural cultures \cite{Neurodelphus,khantanVirtualWhiteMatter2025}.

\begin{table*}[t]
\caption{Comparison of representative SBI platforms and the human brain as computational substrates.}
\label{tab:cl1_neuroplatform_brain}
\centering
\renewcommand{\arraystretch}{1.25}
\setlength{\tabcolsep}{6pt}
\begin{tabular}{m{3.5cm}m{4.2cm}m{4.2cm}m{4.2cm}}
\hline
\textbf{Feature} 
& \textbf{Cortical Labs CL-1} 
& \textbf{FinalSpark Neuroplatform} 
& \textbf{Human Brain} \\
\hline

Biological substrate
& Human iPSC-derived cortical neurons cultured on MEAs
& Human iPSC-derived cortical neurons cultured on MEA
& In-vivo biological neural tissue \\

\hline
Dimensionality
& 2-D and 3-D dissociated neural cultures
& 3-D spheroid cultures
& 3-D highly structured cortical architecture with glial cells \\

\hline
Approx.\ number of neurons
& $\sim 8 \times 10^6$ 
& $\sim 10^4$--$10^6$
& $\sim 8.6 \times 10^{10}$ \\

\hline
Interface technology
& MEA with 59 electrodes
& 4 MEAs with capacity for 4 organoids each. 8 electrodes per organoid.
& Biological sensory and motor pathways \\

\hline
Stimulation modalities
& Electrical stimulation
& Electrical stimulation, limited chemical modulation
& Electrical, chemical, optical, thermal (natural) \\

\hline
Recording modalities
& Extracellular spike recording (MEA)
& Extracellular spike recording (MEA), live camera
& Brain-Machine Interfaces, electroencephalogram \\

\hline
Plasticity and learning
& Activity-dependent synaptic plasticity
& Activity-dependent synaptic plasticity
& Permanent multi-scale plasticity \\

\hline
Stability over time
& Up to 6 months
& Between two and twelve weeks
& Decades \\

\hline
Energy consumption
& Less than $\sim 1~\mathrm{KW}$
& $\sim 20~\mathrm{W}$ 
& $\sim 20~\mathrm{W}$ \\

\hline
Sampling rate
& $25\ \mathrm{kHz}$
& $30\ \mathrm{kHz}$
& - \\

\hline
\end{tabular}
\end{table*}

\section{SBI System Interfaces: The Adaptive Bio-Neural Interaction Architecture}
\label{sec:abnia}

This section introduces a system-level abstraction for SBI system interfaces, termed the \textit{Adaptive Bio-Neural Interaction Architecture} (ABNIA). It presents a framework for the stimulation, sensing, and feedback mechanisms that support SBI platforms. Rather than emphasizing specific biological implementations, ABNIA adopts a system-level perspective, highlighting common architectural components and abstractions that occur across SBI system interfaces. ABNIA is decomposed into four interacting modules:

\begin{itemize}
    \item \textbf{Module I: Encoding and Stimulation Module}: responsible for designing and delivering structured interventions.
    \item \textbf{Module II: Adaptive Biological Neural Substrate}: performs learning-driven, nonlinear information processing.
    \item \textbf{Module III: Observation and Interpretation Module}: captures and decodes neural responses.
    \item \textbf{Module IV: Feedback and Control Module}: adapts stimulation strategies over time.
\end{itemize}

While the presented framework draws analogies with the traditional transmitter-channel-receiver communication scheme, this work does not treat this interface as a communication system per se. This is because SBI systems are not suited for the reliable transmission of symbols, as the neural substrate (which can be paralleled to the channel in communication engineering) is memory-heavy, adaptable and non-stationary \cite{yaronDissociatedNeuronalCultures2025}.

When this work defined SBI system interfaces in Section~\ref{subsec:sbi_definition}, we consider those that use closed-loop schemes, in which stimulation patterns are adapted based on the detected neural activity. This contrasts with open-loop systems, which have no feedback. Building on the testbeds reviewed in Section~\ref{sec:sbi_survey}, we now abstract their common architectural principles.

\begin{figure*}[t]
    \centering
    \includegraphics[width=1.0\linewidth]{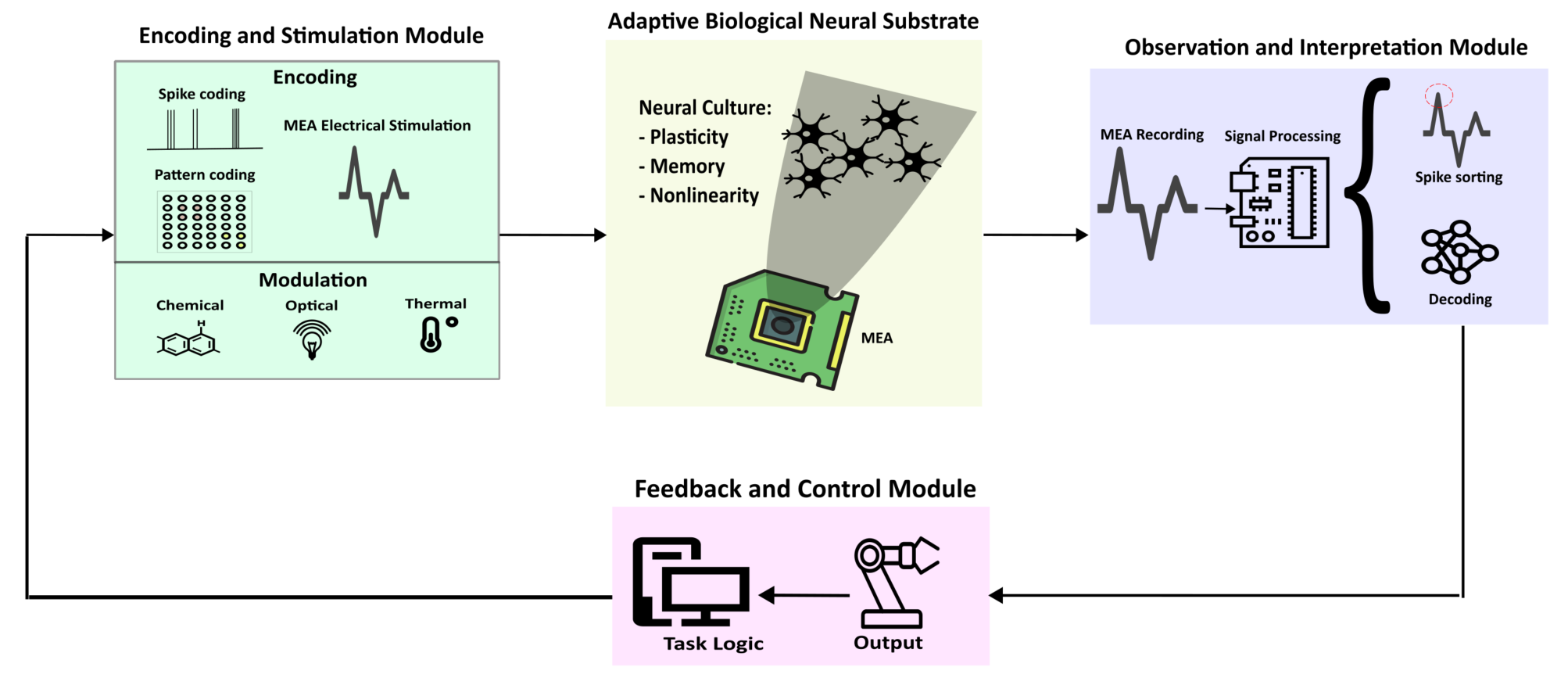}
    \caption{Closed-loop SBI system under the ABNIA framework. Electrical stimulation encodes information through MEAs, while slower modulatory inputs shape channel dynamics. Feedback enables adaptive control of the SBI channel, which exhibits nonlinearity, memory, and plasticity.}
    \label{fig:sbi_comm_pipeline}
\end{figure*}

A generic SBI system framed under the ABNIA pipeline is shown in Fig.~\ref{fig:sbi_comm_pipeline}. Module I, encoding and stimulation, is responsible for generating stimuli in the MEA, encoding spikes, and selecting the most appropriate channels to elicit impulses. Modulatory inputs include chemical perturbations, e.g., neurotransmitters or pharmacological agents, and optical as well as thermal stimuli. Previous interactions act on module II, the adaptive biological neural substrate, the core of the SBI system. Module III, observation and interpretation, comprises both the MEA electrodes that record the response and the post-processing for spike sorting and feature extraction. Then, module IV, feedback and control, provides adaptive stimulation.

An important characteristic of SBI interaction architectures is the presence of multiple interacting time scales, ranging from rapid spike-level dynamics to slower controller updates and long-term synaptic plasticity \cite{shaoOpportunitiesChallengesBrainonaChip2025}. As a result, latency, causality, and timing alignment between stimulation, detection, and control become critical design considerations \cite{duruInvestigationInputoutputRelationship2023}.

\section{Module I: Encoding and Stimulation}
\label{sec:coding_schemes}

Module I, encoding and stimulation in ABNIA, employs various coding and modulation strategies that translate information into stimulation patterns interpretable by the neural substrate. Biological neural substrates are inherently noisy, exhibit strong memory effects, and are subject to biological variability and drift, unlike classical communication channels \cite{huaMicroelectrodeArraysCultured2025}. Appropriate spike encoding schemes should prioritize robustness and interpretability, and this is a crucial research area for harnessing the capabilities of neural cultures.

An important aspect of the coding schemes is that they should account for the fact that neuronal connections are modulated by the input signal in the culture \cite{caporaleSpikeTimingDependentPlasticity2008}. This is mostly due to Spike-Timing-Dependent Plasticity (STDP), which, on a slow timescale, modulates synaptic strength based on the relative timing of spikes. The signals thus shape the SBI neural substrate over time rather than only encoding and transmitting information. A summary of the coding and modulation schemes is presented in Table~\ref{tab:sbi_coding}.

\begin{table*}[t]
\caption{Summary of coding and modulation schemes for ABNIA.}
\label{tab:sbi_coding}
\centering
\renewcommand{\arraystretch}{1.2}
\begin{tabular}{m{2.8cm}m{3.0cm}m{4.2cm}m{3.6cm}m{2.6cm}}
\hline
\textbf{Category} 
& \textbf{Method} 
& \textbf{Definition} 
& \textbf{Typical Usage} 
& \textbf{Examples} \\
\hline

\multirow{2}{*}{\centering Spike-based rate coding}
& Average Firing Rates (AFRs)
& Information is encoded in the mean spike count in a certain period.
& Low bandwidth closed-loop control, responsiveness
& \cite{newmanClosedLoopMultichannelExperimentation2013,guoNeuralCodingSpiking2021} \\
\cline{2-5}

& Tetanic stimulation
& High-frequency stimulation to induce higher firing rates in the population.
& Proof-of-concept activations
& \cite{jimboStrengtheningSynchronizedActivity1998,liNovelRobotSystem2016} \\
\hline

\multirow{3}{=}{\centering \\ Spike-based temporal coding}
& Time-to-First-Spike (TTFS)
& Information is encoded in the latency between the stimulus and the first elicited spike.
& Low-latency, higher bandwidth SBI tasks
& \cite{guoNeuralCodingSpiking2021} \\
\cline{2-5}

& Phase coding
& Information is encoded in the phase of a spike occurrence in relation to an oscillatory activity.
& Energy-efficiency, latency reduction
& \cite{guoNeuralCodingSpiking2021} \\
\cline{2-5}

& Inter-Spike Interval (ISI)
& Information is encoded in the intervals between consecutive spikes in the same network.
& Strengthening synaptic communication
& \cite{bakkumRemovingAIEmbodied2004,bakkumLongTermActivityDependentPlasticity2008,guoNeuralCodingSpiking2021} \\
\hline

\multicolumn{2}{c}{\centering Spatial coding}
& Information is encoded across the spatial distribution of stimulation/recording electrodes in the MEA.
& Multi-class discrimination, sensory encoding, determination of the most suitable electrodes for stimulation or recording
& \cite{olshausenSparseCodingSensory2004,bisioClosedLoopSystemsVitro2019,kaganVitroNeuronsLearn2022,meszarosSpaceTimeNeuron2025} \\
\hline

\multicolumn{2}{c}{\centering Symbolic modulation}
& Discrete symbols are represented by predefined simulation patterns.
& Multi-class discrimination, symbol mapping
& \cite{shahafLearningNetworksCortical2001, augeSurveyEncodingTechniques2021} \\
\hline

\multicolumn{2}{c}{\centering Reservoir-based coding}
& Input signals are given to the substrate to obtain the high-dimensional response from the internal dynamics.
& Computational benchmarking, prediction tasks
& \cite{caiBrainOrganoidReservoir2023} \\
\hline

\multirow{3}{=}{\centering \\ Non-electrical modulation}
& Chemical stimulation
& Modulation of neural activity through neurotransmitters or pharmacological agents.
& Long-term modulation of neuroplasticity, neurodevelopmental, and medical research
& \cite{colombiEffectsAntiepilepticDrugs2013,newmanClosedLoopMultichannelExperimentation2013,parodiVitroElectrophysiologicalDrug2024,parkModulationNeuronalActivity2024,watmuffDrugTreatmentAlters2025} \\
\cline{2-5}

& Optical stimulation
& Light-based modulation of neural activity (typically with optogenetic alterations).
& High-precision stimulation
& \cite{thompsonOpticalStimulationNeurons2014} \\
\cline{2-5}

& Thermal stimulation
& Temperature-based modulation of neurons.
& Specification benchmarking, modulation of neural responsiveness
& \cite{kimThermalEffectsNeurons2022} \\
\hline

\end{tabular}
\end{table*}

One of the simplest yet most widely adopted approaches is \textbf{spike-based rate coding}, in which information is encoded in the firing rate of neurons rather than the timing of individual spikes. This includes Average Firing Rates (AFRs) and pulse train duration \cite{guoNeuralCodingSpiking2021}. These rates can also be modulated by spike amplitude and pulse width. Tetanic stimulation has been widely used in proof-of-concept tests, in which eliciting neural responses was the primary goal \cite{jimboStrengtheningSynchronizedActivity1998}, since it induces specific AFRs. In this scheme, sustained or high-frequency stimulation increases the likelihood of spike generation in a neural culture. This setup is considered relatively robust but offers low bandwidth due to temporal averaging.

A more biologically based approach is \textbf{spike-based temporal coding}, in which information is encoded in the precise timing of spikes, including inter-spike intervals and response latency \cite{bakkumLongTermActivityDependentPlasticity2008,guoNeuralCodingSpiking2021}. Although these theoretically offer better bandwidth, their implementation in SBI is complicated by spike-time jitter and synchronization requirements.

Since MEAs (especially high-density ones) offer different channels in which they can produce stimulation, an important approach is that of \textbf{spatial coding}. This includes different strategies in which information is distributed across the electrode array, considering that neural activity patterns are spatial and present at the population level, rather than appearing exclusively in a single channel \cite{olshausenSparseCodingSensory2004,bisioClosedLoopSystemsVitro2019,kaganVitroNeuronsLearn2022,meszarosSpaceTimeNeuron2025}.

Beyond electrical stimulation, SBI systems can also use non-electrical modulation, such as \textbf{chemical}, \textbf{optical}, and \textbf{thermal} inputs. Rather than encoding discrete symbols, these modalities primarily act on slower time scales and modulate the operating regime of the neural substrate. Pharmacological interventions can alter excitability, synaptic efficacy, or noise levels, effectively changing channel dynamics over longer periods \cite{colombiEffectsAntiepilepticDrugs2013,parodiVitroElectrophysiologicalDrug2024,watmuffDrugTreatmentAlters2025}. Some research has been done on optical stimulation, particularly in optogenetically modified cultures, which enables spatially selective and cell-type activation \cite{thompsonOpticalStimulationNeurons2014,renaultCombiningMicrofluidicsOptogenetics2015}. Thermal modulation can also influence firing rates and network synchrony \cite{kimThermalEffectsNeurons2022}. These forms of modulation are therefore better interpreted as channel-state modulation mechanisms that complement spike-based coding schemes rather than replace them.

\textbf{Symbolic modulation} schemes have also been explored, in which discrete symbols are represented by predefined stimulation patterns. Binary schemes analogous to On-Off-Keying (OOK) \cite{shahafLearningNetworksCortical2001, augeSurveyEncodingTechniques2021} are the most common approach.

Finally, \textbf{reservoir-based coding} does not rely on explicit symbol definitions. The goal of input signals is to generate perturbations in the internal neural dynamics and then interpret the information contained in the system's response. However, these are typically used for computational benchmarking rather than being suitable metrics for ABNIA itself \cite{caiBrainOrganoidReservoir2023}.

In summary, the encoding and stimulation module favors low-order and hybrid coding schemes that prioritize robustness, interpretability, and experimental feasibility. Future research is needed to increase spectral efficiency without sacrificing robustness and provide established protocols for reliable substrate interaction.

\section{Module II: Adaptive Biological Neural Substrate}
\label{sec:channel_characteristics}

In ABNIA, the particular intrinsic dynamics and physical characteristics of the Adaptive Biological Neural Substrate strongly influence its behavior, not only the stimulation itself. These channels are living systems that adapt to their environment, develop over time, and have mortality \cite{talaveraBrainOrganoidComputing2025}. 

Unlike classical communication channels, neural substrates are inherently non-linear, state-dependent, and non-stationary due to ongoing plasticity, spontaneous activity, and feedback-driven modulation \cite{bakkumLongTermActivityDependentPlasticity2008}. As a result, fixed-parametric channel models are generally inapplicable. In addition, several studies suggest that SBI channels operate near critical dynamical regimes, where sensitivity to stimulation and response variability are maximized, thus complicating channel characterization while potentially improving adaptive behavior \cite{habibollahiCriticalDynamicsArise2023}.

Various dispositions have been tested for optimal task-learning performance. Modular cultures are neural cultures within segregated populations and exhibit more reproducible response patterns \cite{levyEnhancementNeuralRepresentation2012, forroModularMicrostructureDesign2018}. This can improve symbol distinguishability and channel stability, at the expense of some flexibility \cite{osakiComplexActivityShortterm2024}.

The substrate-dependent characteristics described above directly influence commonly reported SBI metrics, including decoding accuracy, latency, jitter, and cross-session stability. Given these considerations, performance benchmarking must account for differences between neural substrates, and no substrate can be considered universally optimal for ABNIA.

The adaptive biological neural substrate is best approached through empirical input-output characterization, in which stimulation patterns are related to response features, e.g., firing rates, burst statistics, under a defined protocol, and through state-based descriptions that capture slow dynamics such as excitability, synchrony, or criticality regime. Characterization of the substrate in models has only been done with active inference and generative models \cite{paulSimulatingBiologicalIntelligence2025}, chaotic recurrent neural networks \cite{matteraChaoticRecurrentNeural2025}, Izhikevich models \cite{houbenRoleConnectivityAnisotropies2025}, and liquid state machines \cite{maassRealTimeComputingStable2002}. Developing reproducible and platform-agnostic substrate models that remain valid across cultures, sessions, and substrates remains an open challenge for ABNIA.

\subsection{Noise Sources in Neural Substrates}
\label{subsec:noise}

The adaptive biological neural substrate is subject to multiple noise sources that are distinct from classical additive white Gaussian noise models. These sources include not only intrinsic noise in cells or neuronal synapses, but also interactions with the MEA and environmental conditions \cite{steinNeuronalVariabilityNoise2005}.

The smallest noise source, in terms of size, is the one introduced by stochastic processes in neurons at the biochemical and biophysical levels \cite{faisalNoiseNervousSystem2008}. This includes the opening and closing of ion channels, the diffusion, production, binding, and degradation of proteins and signaling molecules. This noise is amplified when the membrane potential of the neuron is close to the firing threshold \cite{schneidmanIonChannelStochasticity1998}. Other sources of noise at the neuronal membrane include Johnson and shot noise arising from membrane resistance. However, these are insignificant in comparison to the ion channel noise.

On a larger scale, synaptic noise is another source that encompasses the influence of neurotransmitter release and pre- and postsynaptic molecular processes~\cite{ribraultStochasticityMolecularProcesses2011}. This affects the amplitude and timing of postsynaptic action potentials.

At the population level, spontaneous spikes and bursts are also an influential noise source that can complicate the interpretation and decoding of signals \cite{wagenaarExtremelyRichRepertoire2006}. These arise from the culture's developmental stage, the types of neurons used, electrical stimulation in other sections or modules of the neural structure, mechanical sensitivity, or other noise sources previously mentioned. The nature of these bursts is also variable, ranging from tiny to culture-global.

An important consideration is also the influence of the electrodes on the culture \cite{obienRevealingNeuronalFunction2015}. Thermal noise and $1/f$ 
noise, where $f$ denotes the signal frequency arising from the sampling frequency, the electrode impedance, and the MEA circuit, affects the stimulation and recording of neural activity. Thus, these effects must be accounted for in the design and selection of MEAs.

Although not a source of noise per se, the long-term plasticity of neural cultures induces drift and non-stationarity \cite{caporaleSpikeTimingDependentPlasticity2008,eckerLongTermPlasticityInduces2024,eckerLongTermPlasticityInduces2024_1}, which have to be considered when interpreting signals.

\section{Module III: Observation and Interpretation}
\label{sec:decoding_detection}

Decoding schemes in ABNIA aim to obtain task-relevant information from the recorded neural activity. This typically takes the form of spike trains or population-level features. Peri-Stimulus Time Histograms (PSTHs) are commonly used to summarize spike timing statistics relative to stimulation events, serving as temporal features for downstream decoding \cite{tessadoriModularNeuronalAssemblies2012}. The process of determining the presence of spikes and assigning them to specific neurons is called \textit{spike sorting} \cite{buccinoSpikeSortingNew2022}. As mentioned in the previous section, SBI substrates exhibit strong temporal dependencies and biological non-stationarity and variability, which make detection challenging.

\subsection{Spike sorting methods}

\begin{table*}[!t]
\caption{Summary of spike sorting methods.}
\label{tab:spike_sorting}
\centering
\renewcommand{\arraystretch}{1.2}
\begin{tabular}{m{2.8cm}m{3.0cm}m{4.2cm}m{3.8cm}m{2.4cm}}
\hline
\textbf{Pipeline section} 
& \textbf{Method} 
& \textbf{Definition} 
& \textbf{Typical Usage} 
& \textbf{Examples} \\
\hline

\multirow{4}{=}{\centering Raw data filtering}
& Analog casual Infinite Impulse Response (IIR) bandpass filter
& Analog bandpass filtering to eliminate low-frequency drift and high-frequency noise at the electrode prior to digitization.
& Signal conditioning.
& \cite{quianquirogaWhatRealShape2009,park128ChannelFPGABasedRealTime2017} \\
\cline{2-5}

& Digital casual IIR bandpass filter
& Digitally implemented filter for obtaining the frequency components relevant to the spikes in recorded signals. Introduces non-linearities with affect the shape of the spikes.
& Spike isolation.
& \cite{quianquirogaWhatRealShape2009} \\
\cline{2-5}

& Nearly-linear phase IIR filter
& Bandpass filtering with reduced phase distortion to preserve spike waveform shape.
& Spike detection and feature extraction when waveform morphology is relevant.
& \cite{reyPresentFutureSpike2015,park128ChannelFPGABasedRealTime2017} \\
\cline{2-5}

& Zero-phase filter
& Information is obtained from how bursts are composed and distributed in a certain period.
& Offline spike sorting or when latency is not critical, not typically used for closed-loop systems.
& \cite{reyPresentFutureSpike2015} \\
\hline

\multirow{2}{=}{\centering \\ Spike detection}
& Manual threshold
& Fixed voltage threshold to determine the presence of a spike.
& Simple and interpretable spike detection.
& \cite{eytanDynamicsEffectiveTopology2006} \\
\cline{2-5}

& Automatic threshold
& Adaptive thresholding based on signal statistics.
& Robust detection against noise and drift.
&  \cite{quirogaUnsupervisedSpikeDetection2004,reyPresentFutureSpike2015} \\
\hline

\multirow{3}{=}{\centering \\ Feature extraction}
& Template matching
& Extraction of spike features by comparing the detected waveforms with predefined templates.
& Sorting for recordings with limited waveform variability.
& \cite{pachitariuSpikeSortingKilosort42024} \\
\cline{2-5}

& Principal Component Analysis (PCA)
& Dimensionality reduction technique to obtain the most important characteristics of the spike waveforms.
& Classical compression method before clustering. Requires periodic retraining due to drift.
&  \cite{reyPresentFutureSpike2015} \\
\cline{2-5}

& Wavelets
& Time–frequency decomposition of spike waveforms to obtain localized shape features.
& Spike sorting in noisy recordings, or when the waveform shape is different across neurons.
&  \cite{quirogaUnsupervisedSpikeDetection2004,bestelNovelAutomatedSpike2012} \\
\hline

\multirow{6}{=}{\centering \\ Clustering}
& Expectation-Maximization (EM)
& Probabilistic clustering of spike features using mixture models such as Gaussian or t-distributions.
& Probabilistic spike clustering in controlled experimental settings, primarily for offline analysis. Susceptible to drift.
& \cite{bestelNovelAutomatedSpike2012} \\
\cline{2-5}

& ANNs
& Nonlinear classification of spike features using learned representations.
& Automated spike sorting in large-scale or high-density MEA recordings. Requires periodic retraining due to drift.
& \cite{buccinoSpikeSortingNew2022,ohbergNeuralNetworkApproach1996} \\
\cline{2-5}

& Support Vector Machine (SVM)
& Supervised classification of spike features using margin-based decision boundaries.
& Spike classification with limited labeled data and high feature separability.
& \cite{vogelsteinSpikeSortingSupport2004} \\
\cline{2-5}

& k-means
& Unsupervised clustering of spike features based on Euclidean distance minimization.
& Clustering in low-dimensional feature spaces when the number of classes is unknown.
& \cite{caro-martinSpikeSortingBased2018} \\
\cline{2-5}

& Graph-based approaches
& Spike features are represented as nodes in a similarity graph.
& Unsupervised spike clustering in high-dimensional feature recordings, particularly for high-density recordings.
& \cite{pachitariuSpikeSortingKilosort42024} \\
\cline{2-5}

& SNNs
& Highly energy-efficient unsupervised spike recognition.
& Recent technique, aimed at online high-dimensional feature recordings.
& \cite{pokalaFrugalSpikingNeural2025} \\
\hline

\end{tabular}
\end{table*}

Spike sorting is typically set as a multi-stage signal processing pipeline, where the electrode recordings are interpreted to obtain discrete neural events associated to neural units. This pipeline is usually composed of four stages:

\begin{itemize}
    \item \textbf{Raw data filtering}
    \item \textbf{Spike detection}
    \item \textbf{Feature extraction}
    \item \textbf{Clustering}
\end{itemize}

Each stage requires hardware and algorithmic design choices, since these should suit the application for which the spikes are obtained. For SBI systems in particular, spike sorting mechanisms need to consider that closed-loop systems are online, that signals from neural substrates contain noise and drift, and whether the task or decoding mechanism requires that the spikes have a particular shape. Table \ref{tab:spike_sorting} summarizes the different methods for each stage that can be used and in which situations these are suitable.

In the first stage, both analog and digital bandpass filters are used to eliminate low-frequency drift and high-frequency noise, thereby cleaning the signal before spike detection. Causal IIR filters are used for their low latency, but their nonlinearities, introduced by the phase, affect the waveform, which needs to be considered in later stages of spike sorting and decoding. Nearly-linear phase IIR filters can be used if the waveform shape is relevant. Zero-phase filters are not suitable for closed-loop SBI, as they are non-causal and cannot account for future values; they cannot be implemented online and will introduce high latency.

Spike detection in SBI systems is almost always based on automatic thresholds, which use time windows to adaptively detect action potentials, making them robust against noise and drift. Manual thresholds are typically only used for simple, proof-of-concept testbeds.

While not strictly necessary for SBI systems, the feature extraction and clustering stages can be used to obtain a more faithful representation of the neural signals. In some cases, the entire spike sorting pipeline is replaced by algorithms such as ANNs, where the signal is inputted and the information is extracted in a single stage \cite{pizziCulturedHumanNeural2009}. Nevertheless, we mention the main techniques used in these stages, which have also inspired some of the decoding techniques in section \ref{subsec:decoding_techniques}. In particular, feature extraction methods are primarily used to capture the most important characteristics of spike morphology, using popular decomposition and dimensionality reduction techniques such as Principal Component Analysis (PCA) and wavelets. For clustering, supervised and unsupervised methods are used depending on the signal's complexity and the task's particularities.

\subsection{Decoding techniques}
\label{subsec:decoding_techniques}

\begin{table*}[!t]
\caption{Summary of decoding schemes in ABNIA.}
\label{tab:sbi_decoding}
\centering
\renewcommand{\arraystretch}{1.2}
\begin{tabular}{m{2.8cm}m{3.0cm}m{4.2cm}m{3.8cm}m{2.4cm}}
\hline
\textbf{Category} 
& \textbf{Method} 
& \textbf{Definition} 
& \textbf{Typical Usage} 
& \textbf{Examples} \\
\hline

\multirow{3}{=}{\centering Feature-based decoding}
& Spike rate
& Information is obtained from the mean spike count in a certain period.
& Symbol decoding, closed-loop control
& \cite{levyEnhancementNeuralRepresentation2012,isomuraCulturedCorticalNeurons2015,liNovelRobotSystem2016,yangSpikingDynamicsIndividual2025,robbinsGoalDirectedLearningCortical2024} \\
\cline{2-5}

& Spike profile
& Information is obtained from how spikes are distributed in a certain period.
& Symbol decoding
& \cite{tessadoriEncodingStimuliEmbodied2013,tessadoriClosedloopNeuroroboticExperiments2015} \\
\cline{2-5}

& Burst profile
& Information is obtained from how bursts are composed and distributed in a certain period.
& Task learning
& \cite{tessadoriEncodingStimuliEmbodied2013,tessadoriClosedloopNeuroroboticExperiments2015} \\
\hline

\multirow{4}{=}{\centering \\ Classifier-based decoding}
& Linear classifiers
& Neural features are mapped to values that are discriminated by linear decision boundaries.
& Low-latency, low-complexity decoding
& \cite{isomuraCulturedCorticalNeurons2015} \\
\cline{2-5}

& ANNs
& Non-linear mappings are obtained with different types of neural networks.
& Improved accuracy in task learning
& \cite{pizziCulturedHumanNeural2009,aaserMakingCyborgClosedloop2017,buccinoSpikeSortingNew2022} \\
\cline{2-5}

& SVM
& Neural features are classified with decision-boundary optimization.
& Robust classification
& \cite{levyEnhancementNeuralRepresentation2012,liuEncodingTactileStimuli2025} \\
\cline{2-5}

& Unsupervised learning methods
& Neural features are clustered without labels to identify patterns.
& Symbol identification, exploratory decoding
& \cite{demarseNeurallyControlledAnimat2001,levyEnhancementNeuralRepresentation2012} \\
\hline

\multirow{2}{=}{\centering \\ Sequence-aware decoding}
& Functions with memory
& Decoding with the temporal history of the neural features.
& Task-learning, prediction, embodied SBI
& \cite{cozziCodingDecodingInformation2005,demarseAdaptiveFlightControl2005} \\
\cline{2-5}

& Markov processes
& Decoding with state-transition models according to Markov dynamics.
& Temporal pattern decoding, prediction
&  \cite{yangMMTSNNMarkovianDecision2026} \\
\hline

\multicolumn{2}{c}{\centering Filtering}
& Neural features are processed with spatial or temporal filters.
& Signal conditioning
& \cite{ruaroNeurocomputerImageProcessing2005} \\
\hline

\end{tabular}
\end{table*}

A table with the most common decoding schemes is presented in Tab. \ref{tab:sbi_decoding}. The most traditional type of decoding is \textbf{feature-based decoding}, where neural responses are characterized by spike counts, firing and burst rates, and other temporal features within a certain window \cite{tessadoriEncodingStimuliEmbodied2013,levyEnhancementNeuralRepresentation2012}. These are then mapped to symbols or variables using thresholding or decision rules.

In order to attain more complex neural features such as multidimensional or non-linear components, \textbf{classifier-based methods} can be applied. These include linear classifiers \cite{tessadoriEncodingStimuliEmbodied2013}, ANNs \cite{buccinoSpikeSortingNew2022}, or Support Vector Machines (SVMs) \cite{levyEnhancementNeuralRepresentation2012}, often combined with dimensionality-reduction techniques such as Principal Component Analysis (PCA) \cite{buccinoSpikeSortingNew2022,khantanVirtualWhiteMatter2025}. Unsupervised clustering methods, such as k-means, are also commonly used to identify discrete neural response states or to define symbol classes in the absence of explicit labels \cite{khantanVirtualWhiteMatter2025}.

Given the strong memory dependencies in these channels, \textbf{sequence-aware decoding} techniques have been applied \cite{yangSpikingDynamicsIndividual2025}. These account for the temporal correlation introduced by Inter-Symbol Interference (ISI) and recurrent neural dynamics, treating a longer sequence of observations rather than isolated samples.

\section{Module IV: Feedback and Control}
\label{subsec:feedback}

A defining feature of ABNIA is module IV, which closes the feedback loop between the digital and biological components. Unlike open-loop stimulation schemes, closed-loop SBI systems continuously adapt the stimulation pattern based on the observed neural response and the degree to which it deviates from the desired task. This feedback is not solely intended to optimize the given task; it can also support internal regulation and the stabilization of the culture \cite{wagenaarControllingBurstingCortical2005,newmanClosedLoopMultichannelExperimentation2013}.

The works described in Section~\ref{subsec:early_sbi_testbeds} established that neural substrates can be embedded in adaptive control loops, in which task performance metrics directly influence stimulation strategies. More recent SBI platforms have incorporated structured feedback signals to shape network dynamics and induce task-oriented learning. For example, in DishBrain experiments, stimuli were modulated according to task performance in a reinforcement-learning paradigm, thereby serving as a reward signal that guided neural adaptation \cite{kaganVitroNeuronsLearn2022}. Further analysis revealed that feedback-driven learning can shift network dynamics toward critical regimes, enhancing sensitivity and adaptability \cite{habibollahiCriticalDynamicsArise2023}. These findings highlight that feedback not only corrects errors but can actively reshape the dynamical state of the biological substrate.

From the ABNIA perspective, as mentioned in Section~\ref{sec:abnia}, module IV operates across multiple time scales. On fast time scales (milliseconds to seconds), feedback can regulate latency, suppress unstable bursting, and compensate for variability or noise in the neural response \cite{wagenaarControllingBurstingCortical2005,itoChangesNetworkActivity2018}. On intermediate time scales, adaptive decoding and state estimation frameworks—often inspired by control theory and state-space modeling—enable tracking of evolving neural dynamics \cite{brownMultipleNeuralSpike2004}. On slower time scales (minutes to hours), stimulation-dependent synaptic plasticity mechanisms, such as STDP, modify synaptic strengths, thereby embedding learning directly within the substrate \cite{chiappaloneNetworkPlasticityCortical2008}.

As with the other components of ABNIA, feedback in SBI systems also differs fundamentally from classical feedback in communication systems. In Shannon’s abstraction, feedback may improve reliability but does not alter the channel's statistical properties \cite{shannonZeroErrorCapacity1956}. In contrast, biological substrates exhibit plasticity: feedback-driven stimulation can permanently modify connectivity and excitability. Module IV therefore acts not only as a regulator but also as a modulator of substrate dynamics.

Design challenges for this module include ensuring stability under biological variability, preventing runaway excitation, and compensating for drift and non-stationarity. Future research may benefit from formal control-theoretic approaches tailored to adaptive biological systems, including robust control, adaptive control, and reinforcement learning-based strategies.

Within the ABNIA framework, module IV is a central architectural component, mediating between interpretation and stimulation while actively shaping the evolving computational properties of the biological neural substrate.

\subsection{Benchmarking Metrics}
\label{subsec:benchmarking}

Typical metrics for benchmarking decoding performance in ABNIA include classification accuracy, task success rate, and average firing rate, among others, across experiments \cite{bakkumLongTermActivityDependentPlasticity2008,kaganVitroNeuronsLearn2022}. A summary of these metrics across existing testbeds is provided in Table~\ref{tab:sbi_metrics}. Nevertheless, these are not standard across experiments, and a cross-platform benchmark remains an open problem. Given the long-term drift and ever-changing neural response statistics, decoding schemes and metrics for measuring performance remain active areas of research. 

\textbf{Decoding performance} metrics are used to measure how well are symbols or stimuli correctly mapped to their corresponding classes, given a ground truth. These are very related to the clustering methods described in the previous section and Tab. \ref{tab:spike_sorting}, and are relevant for clustering or identification tasks. 

\textbf{Temporal performance} metrics, on the other hand, are focused on the temporal stability and reliability of SBI systems, which is critical in closed-loop and embodied settings.

The suitability of ABNIA systems is best evaluated with its \textbf{task-level performance}, which assesses the functional behavior of the system itself. Metrics such as success rate and task duration until failure are frequently used in neurorobotic environments or control testbeds, to confirm whether the substrate can maintain goal-directed behavior over time. For more signal processing tasks, distribution-based metrics such as the Kullback-Leibler Divergence (KLD) can illustrate the performance of the task.

Metrics regarding \textbf{neural activity} provide insight into the intrinsic dynamics of the biological substrate. Average firing rate, burst rate, and inter-spike interval are very often used to represent the excitability and responsiveness of the network. Another area of interest has been criticality metrics, which are used to measure whether the neural network is operating in near-critical regimes, which offer the optimal dynamic range and processing capability.

Regarding \textbf{learning dynamics} in particular, the temporal evolution of the network when presented with stimuli in order to perform a task is very relevant to address the suitability and robustness of the neural culture. These metrics include the evolution of the task performance of the time, but also information-theoretic ones such as the Mutual Information (MI) or the Response to Stimulus Ratio ($R/S$). Free-energy-based measures can also be used to quantify the learning progress, in particular for prediction tasks.

Finally, centering on \textbf{robustness}, these metrics assess how variable these neural substrates can be. Cross-session stability measures display how consistent the performance is across experiments in the same system, while noise robustness show how sensitive these experiments are when presented with perturbations. These are particularly representative of the reproducibility and standardization capabilities of SBI platforms.

Altogether these performance metric stress that the benchmarking of ABNIA systems is multi-dimensional. Considering the particular aspects of SBI platforms, measuring different types of metrics is necessary to provide point of references to compare them one another.

\begin{table*}[!t]
\caption{Summary of performance metrics used in ABNIA.}
\label{tab:sbi_metrics}
\centering
\renewcommand{\arraystretch}{1.2}
\begin{tabular}{m{2.8cm}m{3.0cm}m{4.2cm}m{3.8cm}m{2.4cm}}
\hline
\textbf{Category} 
& \textbf{Metric} 
& \textbf{Definition} 
& \textbf{Typical Usage} 
& \textbf{Examples} \\
\hline

Decoding performance
& Classification accuracy
& Ratio of correctly decoded symbols or classes, confusion matrix
& Pattern decoding, task inference
& \cite{caiBrainOrganoidComputing2023,caiBrainOrganoidReservoir2023,habibollahiCriticalDynamicsArise2023,liuEncodingTactileStimuli2025} \\
\hline

\multirow{2}{*}{\centering Temporal performance}
& Latency
& Time between stimulus and decoded response
& Closed-loop control, responsiveness
& \cite{novellinoConnectingNeuronsMobile2007,tessadoriModularNeuronalAssemblies2012,zaniniInvestigatingReliabilityEvoked2023} \\
\cline{2-5}

& Jitter
& Variability of response latency
& Stability analysis
& \cite{ruaroNeurocomputerImageProcessing2005,bakkumLongTermActivityDependentPlasticity2008} \\
\hline

\multirow{3}{=}{\centering \\ Task-level\\ performance}
& Task success rate
& Fraction of successful task completions
& Embodied SBI, control tasks
& \cite{novellinoBehaviorsElectricallyStimulated2003,tessadoriModularNeuronalAssemblies2012,tessadoriClosedloopNeuroroboticExperiments2015,khajehnejadBiologicalNeuronsCompete2024,watmuffDrugTreatmentAlters2025} \\
\cline{2-5}

& Task duration until failure
& Experiment time until failure
& Embodied SBI, control tasks
& \cite{novellinoBehaviorsElectricallyStimulated2003,tessadoriModularNeuronalAssemblies2012,tessadoriClosedloopNeuroroboticExperiments2015,itoChangesNetworkActivity2018,khajehnejadBiologicalNeuronsCompete2024,robbinsGoalDirectedLearningCortical2024,hofmannControllingCyberPhysicalSystem2025,watmuffDrugTreatmentAlters2025} \\
\cline{2-5}

& Kullback-Leibler Divergence (KLD)
& Comparison of probability distributions
& Signal processing, clustering and differentiation tasks
& \cite{isomuraCulturedCorticalNeurons2015} \\
\hline

\multirow{4}{=}{\centering \\ Neural activity}
& Average Firing Rate (AFR)
& Mean spike rate over response window
& Channel characterization and response identification
& \cite{demarseAdaptiveFlightControl2005,ruaroNeurocomputerImageProcessing2005,novellinoConnectingNeuronsMobile2007,tessadoriModularNeuronalAssemblies2012, itoChangesNetworkActivity2018,kaganVitroNeuronsLearn2022,zaniniInvestigatingReliabilityEvoked2023,jordanOpenRemotelyAccessible2024,liuEncodingTactileStimuli2025,watmuffDrugTreatmentAlters2025} \\
\cline{2-5}

& Burst rate
& Frequency of burst events
& Network dynamics analysis
& \cite{martinoiaEmbodiedVitroElectrophysiology2004,novellinoConnectingNeuronsMobile2007,habibollahiCriticalDynamicsArise2023,zaniniInvestigatingReliabilityEvoked2023,watmuffDrugTreatmentAlters2025} \\
\cline{2-5}

& Inter-Spike Interval
& Time between spikes
& Network dynamics analysis
& \cite{watmuffDrugTreatmentAlters2025} \\
\cline{2-5}

& Criticality
& Indicators of near-critical network dynamics
& Channel state characterization
& \cite{habibollahiCriticalDynamicsArise2023,watmuffDrugTreatmentAlters2025} \\
\hline

\multirow{4}{=}{\vspace{-1cm} Learning dynamics}
& Free-energy measures
& Variational Free Energy or entropy-related values
& Network learning, prediction and adaptation measurement
& \cite{isomuraCulturedCorticalNeurons2015,fristonFreeenergyPrincipleUnified2010,kaganVitroNeuronsLearn2022} \\
\cline{2-5}

& Task performance over time
& Goal-related metric evolution across time
& Plasticity and adaptation
& \cite{bakkumMEARTSemilivingArtist2007,tessadoriModularNeuronalAssemblies2012,liNovelRobotSystem2016,caiBrainOrganoidReservoir2023} \\
\cline{2-5}

& Mutual Information (MI)
& Shannon entropy comparison between neuronal activity and a future stimulus
& Prediction quantification
& \cite{lambertiPredictionCulturedCortical2023} \\
\cline{2-5}

& Response to Stimulus Ratio ($R/S$)
& Moving average of spike rate within a post stimulus interval
& Network learning and adaptation measurement
& \cite{pimashkinAdaptiveEnhancementLearning2013} \\
\hline

\multirow{2}{=}{\\ Robustness}
& Cross-session stability
& Performance variation across sessions
& Reproducibility assessment
& \cite{itoChangesNetworkActivity2018,habibollahiCriticalDynamicsArise2023,robbinsGoalDirectedLearningCortical2024} \\
\cline{2-5}

& Noise robustness
& Variation in performance with different magnitudes of noise
& Channel characterization, reproducibility assessment
& \cite{liuEncodingTactileStimuli2025} \\
\hline

\end{tabular}
\end{table*}

\section{Challenges and Open Questions}
\label{sec:future_work}

Despite recent progress in SBI platforms and interaction interfaces, several open challenges remain before ABNIA can be systematically engineered, benchmarked, and deployed. These challenges span biomedical considerations, system integration, and standardization, and reflect the inherent tension between biological adaptability and interpretation reliability.

\subsection{Current Limitations}
\label{subsec:limitations}

Despite encouraging proof-of-concept demonstrations, current ABNIA systems remain constrained by several fundamental limitations. First, most existing implementations operate under tightly controlled experimental conditions and are optimized for specific tasks or platforms. As a result, reported interaction architectures and performance metrics are often difficult to generalize beyond the original experimental setup.

Second, SBI systems exhibit pronounced non-stationarity due to ongoing plasticity, spontaneous activity, and long-term drift. While adaptive and closed-loop strategies can partially compensate for these effects, they also complicate substrate characterization, performance evaluation, and repeatability. In many cases, system performance depends on stimulation history, cultural age, and experimental timing, thereby limiting the applicability of static coding or decoding schemes.

Third, the lack of standardized benchmarking tasks and evaluation protocols complicates reliable comparison across SBI systems. Metrics are frequently chosen based on task convenience or biological observability rather than communication-theoretic considerations, and negative or unstable experimental results are rarely reported. This introduces selection bias and obscures critical failure modes for system-level engineering.

Finally, practical constraints related to biological maintenance, culture viability, and interface complexity limit the duration of experiments, scalability, and deployment. These constraints currently limit ABNIA systems to laboratory environments, preventing their systematic integration into larger infrastructures. Nevertheless, cloud-based applications can make up for the lack of infrastructure. 

Addressing these limitations is a prerequisite for developing reproducible, scalable, and interoperable ABNIA systems. The following subsections outline key research directions to address these challenges.

\subsection{Future Research Pathways}
\label{subsec:future_paths}

The limitations described in the previous section are merely considerations to keep in mind when moving forward with new applications or ABNIA schemes. There are many research lines with promise for leveraging many of the benefits SBI offers.

\subsubsection{SBI as a Hybrid System Co-Processor}
\label{subsubsec:coprocessor}

Viewing SBI as a co-processor highlights its role as a complementary biological module embedded within a larger digital system. This is discussed in Section~\ref{sec:abnia} and illustrated in Fig.~\ref{fig:sbi_comm_pipeline}. In this configuration, the SBI component provides non-linear signal processing, temporal integration, and adaptive memory through its intrinsic neural dynamics. On the other hand, conventional digital components handle orchestration, deterministic control, and performance evaluation. This co-processing paradigm raises fundamental questions regarding task partitioning, interface design, and system-level integration.

The operation of ABNIA shares conceptual similarities with \textit{semantic communication} frameworks, in which the objective is not the faithful transmission of symbols, but the conveyance of task-relevant information under strong noise and variability constraints \cite{lanWhatSemanticCommunication2021}. An open research direction is the explicit integration of semantic communication principles into SBI systems, potentially enabling task-aware and task-oriented encoding, adaptive semantic compression, and robustness to biological variability \cite{xieDeepLearningEnabled2021}. This makes SBI a compelling candidate for hybrid intelligence architectures envisioned in beyond-5G systems.

Open challenges include determining which signal-processing or task-learning functions are best delegated to SBI modules, how to integrate biological co-processors with digital controllers, and how to balance learning-driven adaptation with the need for predictable and stable system behavior. Addressing these questions will be essential for leveraging SBI as a reliable and interoperable component within hybrid bio-digital systems.

\subsubsection{Biomedical and Neuroengineering Contexts}
\label{subsubsec:biomedical}

From a biomedical and neuroengineering perspective, SBI systems raise important questions about how biological variability and pathology influence information transmission and control \cite{mencattiniAssembloidLearningOpportunities2024}. Neural cultures derived from different cell types, developmental stages, or disease models exhibit different channel characteristics, including altered noise levels, response latency, and plasticity dynamics \cite{ajongboloBrainOrganoidsAssembloids2025}. Similarly, pharmacological interventions can modulate excitability and synaptic connectivity, thereby perturbing channels and affecting decoding reliability and closed-loop stability \cite{nichollsElectricalStimulationStem2025}.

In particular, neurodegenerative diseases such as Parkinson's and Alzheimer's disease, together with neural disorders such as multiple sclerosis and epilepsy, and psychiatric disorders such as schizophrenia, bipolar disorder, and depression, involve progressive disruptions of neural network dynamics \cite{yadavBrainOrganoidsTiny2021}. These can potentially be studied and modulated using in-vitro SBI systems \cite{colombiEffectsAntiepilepticDrugs2013,newmanClosedLoopMultichannelExperimentation2013,sharfFunctionalNeuronalCircuitry2022,parodiVitroElectrophysiologicalDrug2024,watmuffDrugTreatmentAlters2025}. The insights produced from the SBI learning for task completion can also be useful to improve the performance of BMIs, as sensory-motor functions or human cognition can be extrapolated from these systems \cite{bonifaziVitroLargescaleExperimental2013,newmanClosedLoopMultichannelExperimentation2013}.

Understanding how these factors affect ABNIA performance remains an open research direction, while some work has been done \cite{watmuffDrugTreatmentAlters2025,sunBioengineeringInnovationsNeural2025,birteleModellingHumanBrain2025}. Systematic investigation of diseased or chemically perturbed neural substrates could enable the use of ABNIA metrics as functional indicators of neural health, while also informing the design of robust system architectures that generalize across biological conditions and structures.

\subsubsection{Standardization and Reproducibility}
\label{subsubsec:standardization_reproducibility}

Current SBI studies often rely on platform-specific interfaces, experimental protocols, and evaluation criteria, limiting cross-platform comparison and cumulative progress. Establishing standardized system abstractions, benchmarking tasks, and reporting practices is therefore essential for the maturation of the field. The aforementioned state-of-the-art platforms represent a significant advance in enabling reproducible interfaces and experiments, but the SBI field is still in its early stages of developing industry-wide standards.

Given neuronal mortality \cite{talaveraBrainOrganoidComputing2025} and intrinsic variability between cultures, ensuring the reproducibility of robust substrate models is essential for their implementation in future communication networks. In this context, an important concept is \textit{Transfer Learning}, a pillar of beyond-5G communications \cite{wangTransferLearningPromotes2021}. Not only is this relevant for reproducing specific setups, but it can also be applied to more efficient learning of new tasks. 

Another approach that could improve the reproducibility of ABNIA is to model it for use in simulations. While recently there have been some developments, more options are needed to provide reliable Digital Twins.

Open questions include how to define platform-agnostic interaction protocols, how to characterize the effective information capacity of adaptive biological channels, and how to ensure transferability of ABNIA schemes across SBI substrates and implementations. Progress in this direction would enable systematic benchmarking, facilitate reproducibility between laboratories, and support the development of interoperable SBI communication systems.

\section{Conclusion} 
\label{sec:conclusion}

This work presents a comprehensive survey of SBI from a system-interaction perspective. We have provided a unified definition of SBI, reviewed its historical development and current platforms, and introduced ABNIA, a framework to characterize SBI interfaces. In particular, this survey has detailed the encoding and decoding schemes employed in SBI systems, the unique properties of SBI neural substrates, and the architectural components required for closed-loop interaction between the biological and digital devices. Finally, we have outlined open challenges related to standardization, benchmarking, and protocol design that must be addressed for SBI to mature as an engineering field.

The emergence of SBI represents a paradigm shift in how intelligence and computation can be conceived, moving beyond purely in-silico implementations toward hybrid systems that directly exploit living biological substrates. Although SBI remains in an early stage and is currently confined to research environments, it exhibits strong potential for ultra–energy-efficient computation, rapid learning from limited data, and adaptive information processing under noisy conditions. By framing SBI within ABNIA, this survey aims to provide a foundation for future research at the intersection of biological computation, communication engineering, and beyond-5G and next-generation network architectures.

\bibliographystyle{IEEEtran}
\bibliography{references}

\vfill

\end{document}